%% file: main.tex
\documentclass[journal]{IEEEtran}
\usepackage{amsmath,amsfonts}
\usepackage[linesnumbered,ruled,vlined]{algorithm2e}
\usepackage{array}
\usepackage[caption=false,font=footnotesize,labelfont=rm,textfont=rm]{subfig}
\usepackage{textcomp}
\usepackage{stfloats}
\usepackage{url}
\usepackage{verbatim}
\usepackage{graphicx}
\usepackage{cite}
\usepackage{svg}
\graphicspath{{figure/}}
\usepackage{threeparttable}
\usepackage{multirow}

\begin{document}
\title{A High-Throughput FPGA Accelerator for \\ Lightweight CNNs With Balanced Dataflow}

\author{\IEEEauthorblockN{Zhiyuan Zhao, Yihao Chen, Pengcheng Feng, Jixing Li, Gang Chen, Rongxuan Shen, Huaxiang Lu}
\thanks{
This work was supported by the CAS Strategic Leading Science and Technology Project XDB44000000. (Corresponding author: Gang Chen.)

Zhiyuan Zhao is with the School of Microelectronics, University of Science and Technology of China, Hefei 230026, China, and also with the Institute of Semiconductors, Chinese Academy of Sciences, Beijing 100083, China (e-mail:zhiyuan-zhao@outlook.com).
Yihao Chen, Pengcheng Feng, Jixing Li, Gang Chen, Rongxuan Shen, and Huaxiang Lu are with the Institute of Semiconductors, Chinese Academy of Sciences, Beijing 100083, China, and also with the University of Chinese Academy of Sciences, Beijing 100089, China (e-mail:chengang08@semi.ac.cn).}

}



\maketitle

\input{chapters/abstract.tex}
\input{chapters/Introduction.tex}

\input{chapters/Background.tex}

\input{chapters/Architecture.tex}
\input{chapters/Dataflow.tex}

\input{chapters/Allocation.tex}
\input{chapters/Experiments.tex}
\input{chapters/Conclusion.tex}

\bibliographystyle{IEEEtran}
\bibliography{reference}

\newpage






\end{document}

%% file: chapters/abstract.tex
\begin{abstract}
FPGA accelerators for lightweight convolutional neural networks (LWCNNs) have recently attracted significant attention. 
Most existing LWCNN accelerators focus on single-Computing-Engine (CE) architecture with local optimization.
However,  these designs typically suffer from high on-chip/off-chip memory overhead and low computational efficiency due to their layer-by-layer dataflow and unified resource mapping mechanisms.
To tackle these issues, a novel multi-CE-based accelerator with balanced dataflow is proposed to efficiently accelerate LWCNN through memory-oriented and computing-oriented optimizations.
Firstly, a streaming architecture with hybrid CEs is designed to minimize off-chip memory access while maintaining a low cost of on-chip buffer size.
Secondly, a balanced dataflow strategy is introduced for streaming architectures to enhance computational efficiency by improving efficient resource mapping and mitigating data congestion.
Furthermore, a resource-aware memory and parallelism allocation methodology is proposed, based on a performance model, to achieve better performance and scalability.
The proposed accelerator is evaluated on Xilinx ZC706 platform using MobileNetV2 and ShuffleNetV2.
Implementation results demonstrate that the proposed accelerator can save up to 68.3\% of on-chip memory size with reduced off-chip memory access compared to the reference design. 
It achieves an impressive performance of up to 2092.4 FPS and a state-of-the-art MAC efficiency of up to 94.58\%, while maintaining a high DSP utilization of 95\%, thus significantly outperforming current LWCNN accelerators.

\end{abstract}


%% file: chapters/Introduction.tex
\section{Introduction}


The acceleration of lightweight neural convolutional network (LWCNN) on FPGA has drawn tremendous research attention in recent years\cite{bai2018cnn,yu2020light, wu2019high,wu2021flexible,knapheide2020high, li2021dynamic}. 
FPGAs have been widely adopted for CNN acceleration thanks to their high configurability, excellent power efficiency, and robust computational capabilities.
Meanwhile, LWCNNs reduce both computational complexity and memory usage while maintaining acceptable accuracy, making them ideal for resource-constrained scenarios.
FPGA-based LWCNN accelerators promise higher performance and power efficiency,
which are crucial for high-speed applications\cite{watanabe2014architectures} with high frame rate demands or energy-efficient edge devices\cite{yu2020light}, such as self-driving automobiles, high-speed robotics, and drones.

\begin{figure}[!t]
\centerline{\includegraphics[width=\linewidth]{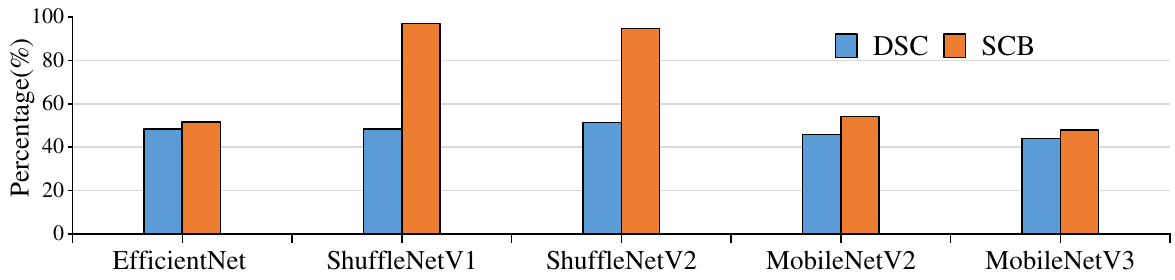}}
    \caption{Percentage of DSC and SCB structures in major LWCNNs.}
    \label{fig:percentage}
    \vspace{-0.3cm}
\end{figure}

Depthwise separable convolutions (DSCs) and skip-connection blocks (SCBs) play an essential role in realizing compact and efficient LWCNN models.
DSCs decompose a standard convolution (STC) into a depthwise convolution (DWC) followed by a pointwise convolution (PWC),  
significantly reducing the number of computations and parameters.
SCBs\cite{myfirst2024}, derived from ResNet\cite{he2016deep}, utilize short connections to mitigate the accuracy degradation problem during the training of deep networks.
As shown in Fig.~\ref{fig:percentage}, DSCs and SCBs account for a significant percentage of the model structure.
Therefore, computational optimization for these components is critical for accelerator design.
However, 
both structures reduce data reuse opportunities and operational intensity in LWCNNs, 
posing challenges for achieving anticipated performance improvements on conventional CNN accelerators\cite{yu2020light, wu2021flexible}.

\IEEEpubidadjcol
To support LWCNNs efficiently, a number of dedicated hardware architectures have been developed. 
Most of these architectures focus on single-Computing-Engine (CE) designs with local optimizations to decrease data communication or improve PE efficiency.
References \cite{bai2018cnn,yu2020light} slice and map DWC layers into a unified convolution CE to achieve higher computational parallelism.  
To reduce intermediate FM transfers, a dedicated CE is deployed in \cite{wu2019high,wu2021flexible,knapheide2020high} to process DWC layers separately.
This approach allows DWC and PWC layers to operate in parallel within a pipeline, resulting in lower latency.
In contrast, \cite{xie2019fast} makes use of partial fusion dataflow to alleviate the bandwidth burden of SCB,
while \cite{yan2021fpga} employs an additional input buffer for SCB inputs to eliminate data transmission between on-chip and off-chip memory.
Despite these advancements, all these designs essentially execute LWCNNs with a single CE and a fixed data reuse scheme.
Their layer-by-layer dataflow demands either heavy off-chip memory traffic due to frequent FM and weight access or a large on-chip buffer footprint\cite{DepFiN23}, which degrades system efficiency.
Meanwhile, the unified computing resource mapping mechanism and fixed data reuse scheme in a single CE may not fit well for the bandwidth or computing resource requirements of each layer \cite{maximizing17, highutilization22, yolotiny24}, given the substantial differences between LWCNN layers in terms of data shape and the number of operations. 
This normally leads to low computational efficiency during calculation (typically less than 50\%) and accordingly produces a significant gap between actual and theoretical throughput. 
Furthermore, most of the aforementioned works still face a problem of insufficient utilization of available computing resources. 
The DSP utilization (normally less than 70\%) on deployed FPGA is not considered in these works, 
which results in wasted DSPs and unnecessary performance loss on the target platform.

To address these issues,
a novel streaming architecture with hybrid CEs is first proposed to relieve considerable FM bandwidth stress caused by DSCs and SCBs.
On the observation of local data dependency between contiguous layers and the distributions of FM and weight sizes in LWCNNs,
our accelerator fully reuses intermediate results between CEs to minimize DRAM access for FMs,
while feature map reused CE (FRCE) and weight reused CE (WRCE) are deployed for shallow and deep layers, respectively, to decrease on-chip memory overhead and weight access.
This approach reduces on-chip memory occupation by 56.67\% compared to the state-of-the-art streaming accelerator in \cite{jiang2023high}, with modest off-chip traffic and the same LWCNN model.
Then, a balanced dataflow strategy is designed to improve the computing efficiency of the multi-CE-based accelerator from two perspectives:
a fine-grained parallel mechanism (FGPM) to enlarge parallel space and thus enable efficient resource mapping at a theoretical level, 
and a dataflow-oriented line buffer scheme to relieve data congestion at an implementation level.
This results in up to a 5.8$\times$ improvement in MAC efficiency compared to previous single-CE-based architectures with a unified computing resource mapping scheme.
Furthermore,
since DSP utilization and CE types can be flexibly adjusted based on the highly parameterized CE design,  
a resource allocation methodology is proposed to explore the design space and leverage limited computing and memory resources on the target platform.
The proposed architecture is evaluated with four representative LWCNNs and is further implemented on MobileNetV2 and ShuffleNetV2 in accordance with the described techniques.
A high performance of 2092.4 FPS and an almost full-load MAC efficiency of 94.58\% are achieved, which outperforms the previous works.

The main contributions of this paper are given as follows:
\begin{itemize}
\item A streaming architecture with hybrid CEs is proposed to minimize on-chip memory overhead as well as off-chip access, where the FRCE targets optimizing FM buffers for shallow layers, while the WRCE aims at maximizing off-chip weight reuse for deep layers.
\item A balanced dataflow strategy, including a fine-grained parallel mechanism and a dataflow-oriented line buffer scheme, is proposed to enhance computing efficiency through efficient resource mapping and data congestion mitigation. 
\item Based on a performance model, a resource-aware memory and parallelism allocation methodology is proposed to improve the scalability of the accelerator to further enhance system performance.
\end{itemize}

The rest of this paper is organized as follows. 
Section II introduces the background and motivation. 
Section III presents the accelerator with hybrid CEs. 
In Section IV, a balanced dataflow strategy based on the proposed architecture is discussed. 
Section V outlines the resource allocation algorithm. 
Section VI shows the experimental results.
Finally, the conclusion is drawn in Section VII.

%% file: chapters/Background.tex
\section{Background \& Related Work}
\subsection{LWCNNs with DSCs \& SCBs}\label{sec:lwcnn}
\begin{figure}[!t]
    \centering
    \subfloat[]{\includegraphics[width=0.8\linewidth]{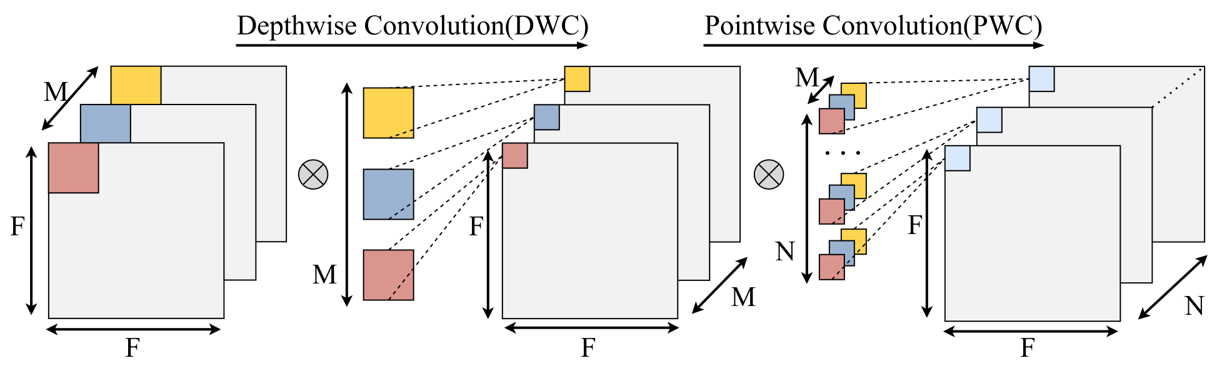}\label{fig:dsc}}\\%
    \subfloat[]{\includegraphics[width=0.8\linewidth]{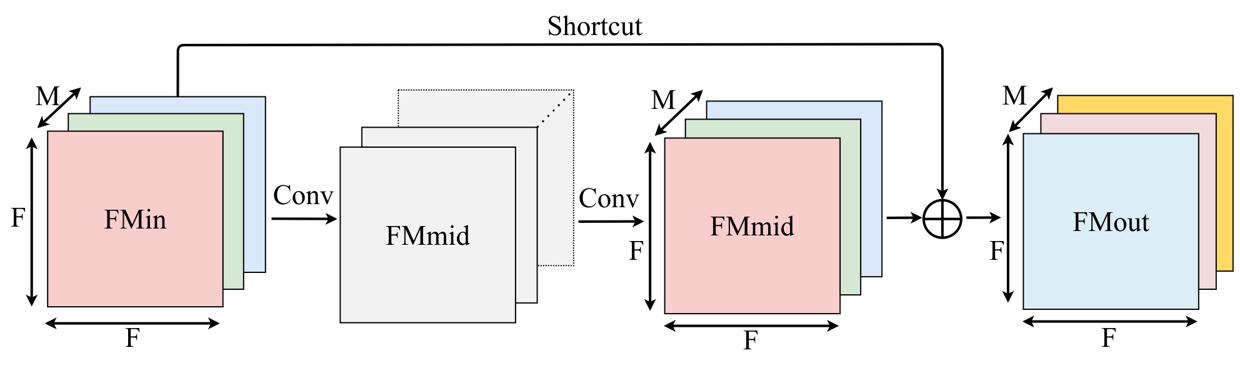}\label{fig:blo}}%
     \caption{Low computational density operations in LWCNNs. (a) Depthwise Separable Convolution(DSC); (b) Skip-Connection Block(SCB).}
    \label{fig:dsc blo}
\end{figure}

STC layers generally occupy the largest amount of operations in traditional CNNs (e.g., VGG\cite{simonyan2014very}).
To reduce model complexity while maintaining acceptable accuracy in LWCNNs, DSC,  as shown in Fig.~\ref{fig:dsc blo}\subref{fig:dsc}, decomposes STC into a DWC and a PWC, where DWC only performs convolution operations within separate channels, and PWC is a kind of STC with a kernel size of 1$\times$1 to realize information interaction and extraction across channels.
Additionally, as the depth of modern networks increases, SCBs, inspired by ResNet\cite{he2016deep}, are also deployed for better final accuracy in many representative LWCNNs.   
A classic SCB structure consists of several convolutional layers with one shortcut branch, as shown in Fig.~\ref{fig:dsc blo}\subref{fig:blo}.
Unlike conventional layers, shortcut branches add only a limited number of addition operations without extra weights, which has a minor effect on model computational complexity and size.


However, both DSC and SCB involve additional memory access for FMs, which may degrade hardware performance.
For better illustration, the number of MAC operations (O) and FM memory access cost (A) are introduced to reveal the relationship between memory access bandwidth requirements and computing resources in different structures.
Without loss of generality, 
it is assumed that the stride is one and padding operations are included.
The kernel size is $K \times K$, and the FM size is $F \times F$ with input and output channels $M$, and $N$, respectively. 
For SCBs, the number of output channels is equal to the input.
The MAC numbers for STC, DSC, and SCB are then calculated as follows:
\begin{align}
O_{S T C}&=F^2 \times K^2 \times M \times N   \\
O_{D S C} & =O_{D W C}+O_{P W C} =F^2 \times M \times\left(K^2+N\right)  \\
O_{SCB}&=\frac{M \times F^{2}}{2} \label{equ:O_SCB}
\end{align}
Note that the number of MACs needs to be halved in Equation~\eqref{equ:O_SCB} since there are only addition operations in SCB.
Then, the FM memory access costs can be calculated as follows:
\begin{align}
A_{S T C}&=F^2 \times(M+N) \\
A_{D S C} & =F^2 \times(3M+N)\\
A_{SCB}&=M_{in} + M_{mid} + M_{out} =3 \times M \times F^{2}
\end{align}
Consequently, the ratios of DSC and SCB to STC in terms of MAC operations ($RO_{DSC}$, $RO_{SCB}$) and FM memory access ($RA_{DSC}$, $RA_{SCB}$) are computed as follows:
\begin{align}
RA_{DSC} & =\frac{A_{DSC}}{A_{S T C}} = 1+\frac{2 M}{M+N} \\
RO_{DSC} & =\frac{O_{D S C}}{O_{S T C}} = \frac{1}{N}+\frac{1}{K^2}  \\
RA_{SCB} & =\frac{A_{SCB}}{A_{S T C}} = \frac{3M}{M+N}  \\
RO_{SCB} & =\frac{O_{SCB}}{O_{S T C}}  =\frac{1}{2N\times{K^2}} 
\end{align}

Based on the above equations, DSC reduces operations by nearly $K^2$ times compared to STC but increases FM access by about one time due to the additional intermediate results transfers. 
Similarly, SCB also incurs numerous memory access costs that are highly disproportionate to the computational load. 
Compared with traditional CNNs, LWCNNs have higher memory access intensity with high bandwidth requirements.
This leads to either significant data transfers between off-chip memory and the accelerator or considerable on-chip memory overhead in the traditional layer-by-layer computational mode\cite{li2021dynamic, yu2020light, yan2021fpga, chen2019eyeriss}. Therefore, it is essential to develop an architecture optimized for memory access, especially FM access, to fully exploit the low computational demands of LWCNNs for higher performance.
\subsection{Distribution of FMs and Weights in LWCNNs}\label{sec:distribution}
\begin{figure}[!t]
    \centering
    \subfloat[]{\includegraphics[width=\linewidth]{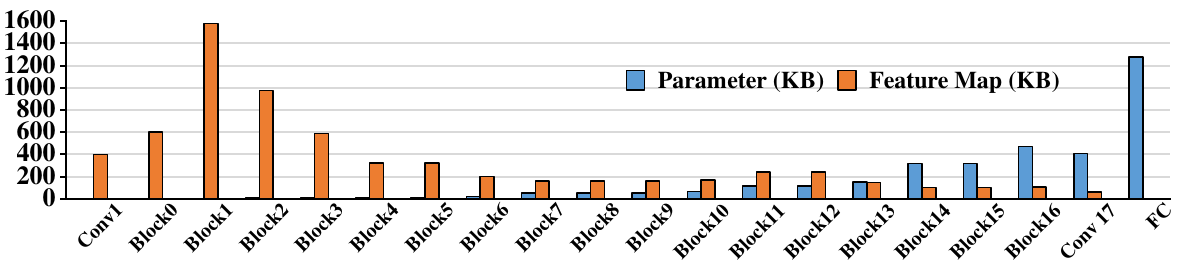}\label{fig:mobilenetv2fmw}}\\%
    \subfloat[]{\includegraphics[width=\linewidth]{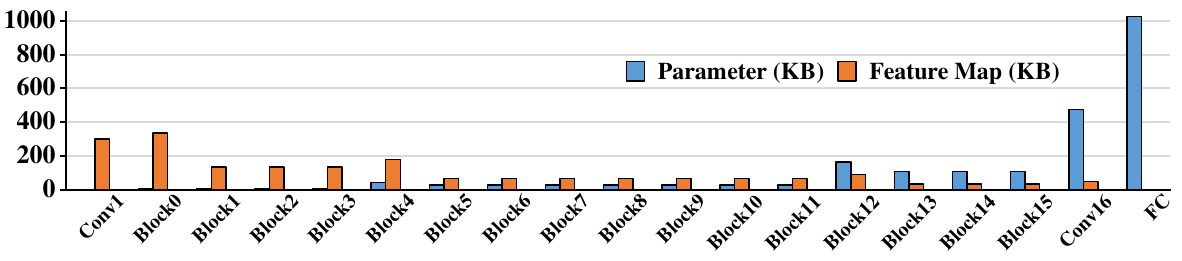}\label{fig:shufflenetv2fmw}}%
     \caption{Memory requirements for FMs and Weights in LWCNNs. The data for each block is the sum of all the layers within it. (a) MobileNetV2; (b) ShuffleNetV2.}
    \label{fig:data distribution}
\end{figure}


CNNs generally contain tens to hundreds of serially stacked computational layers. 
After an image is input into the network, the FMs generated by each layer capture increasingly complex and abstract features as the data propagates from shallow layers to deep layers. 
During this process, along with the reduction of the spatial resolution of FMs, the output channels are gradually increased to compensate for the information loss \cite{726791}. 
This leads to a common phenomenon in both CNNs and LWCNNs: the number of parameters and weights varies from layer to layer but exhibits a similar distribution among networks,
where the FM size is larger than the weight size in shallow layers, and conversely, the deep layers generate fewer FMs but consume massive parameters. 
As illustrated in Fig.~\ref{fig:data distribution}(a), with a 224 $\times$ 224 input size and 8-bit precision, the first STC layer in MobileNetV2 produces 400KB of FMs while using merely 896 parameters. In contrast, the weight size in the last PWC layer is almost 26 times the size of input activations. 
Similarly, Fig.~\ref{fig:data distribution}(b) shows that the ShuffleNetV2 maintains the same FM and weight distributions with fewer parameters.

For hardware design, different FM and weight distributions lead to distinct memory footprints and data reuse rates in each layer,
which made it inappropriate to employ a common computational pattern for the entire model since a fixed pattern may incur redundant memory costs to accommodate the largest data size among all layers.
Several previous designs\cite{SmartShuttle18,azizimazreah2019shortcut,nguyen2021layer, ShortcutFusion22} have adopted hybrid data reuse patterns to minimize memory overhead.
However, these works primarily aim for CNN applications with poor support for lightweight models.
For example, ShortcutFusion in \cite{ShortcutFusion22}, which supports two kinds of weight reuse schemes, achieves merely 19.37\% MAC efficiency when performing DSC-based EfficientNet due to its lower computational density.
In \cite{li2021dynamic}, two types of DSC computation orders are employed for on-chip memory reduction.
The inter-channel first computing order is applied in shallow layers to save FM storage size, while the intra-channel first computing order is adopted in deep layers to save weight data footprint.
Although this approach effectively reduces the on-chip memory occupation for DSC computation, lower computing efficiency and high power consumption are caused by their complicated PE structure and inefficient resource mapping.
In this design, two types of reuse patterns are also exploited to optimize both on-chip and off-chip memory overhead for the proposed streaming architecture.
\parskip=0pt

\subsection{Related Works About LWCNN Accelerators}
Many LWCNN accelerators have been proposed to support LWCNNs effectively.
Some of these accelerators perform all kinds of convolutions within a unified CE \cite{bai2018cnn,yu2020light,yan2021fpga, HighspeedLowcost20}.
They generally optimize the processing of DWC and STC layers to increase resource utilization.  
For example, 
Light-OPU \cite{yu2020light} uses an additional parallel dimension for DWC layers to ensure sufficient parallel space.
Nevertheless, low computational efficiency is caused by frequent transfers of massive intermediate activations between external memory and the unified CE.
\parskip=0pt

To reduce the off-chip access burden for FMs, a dedicated CE is developed for DWC processing in a separated CE architecture \cite{wu2019high,wu2021flexible,knapheide2020high}.
By fusing PWC and DWC layers, the intermediate FMs generated by the PWC CE can be directly transferred to the DWC CE without off-chip transmission. 
In this way, the PWC CE and DWC CE can be processed in a pipeline mode to relieve the bandwidth stress for DWC layers.
However, due to the large difference in computational workloads between the DWC and PWC layers, the DWC CE with fixed computing resources cannot always ensure high run-time efficiency, especially for non-DSC structures where the DWC CE remains completely unused, resulting in a waste of hardware resources.
\parskip=0pt

To optimize the bandwidth problem in the SCB structure, 
\cite{yan2021fpga} introduce an additional buffer for input data storage and eliminate the need for shortcut data transmission,
thereby consuming considerable on-chip memory.
Moreover, partial fusion dataflow has been studied in \cite{xie2019fast},
where input FMs of the first layer are split into several tiles, and each tile is processed across all layers of the SCB with a fusion approach. 
This avoids FM access for middle layers and shortcut connections. 
However, repetitive weight access is caused by the low reuse rate of weight data, which can only be conducted within the tile being processed.  
Ideally, a fusion dataflow scheme should ensure all off-chip data is accessed only once.
\parskip=0pt

There are existing works that present a multi-CE architecture for LWCNNs\cite{jiang2023high, ding2019designing, UltraEnergyEfficient22, myfirst2024}, where each layer is processed by a dedicated CE to enable high utilization of computing resources.  
For example, \cite{UltraEnergyEfficient22} employs SkyNet with all weights stored in BRAM. 
However, due to limited on-chip memory capacity, this approach cannot scale effectively for deeper networks. 
\cite{jiang2023high} implements a full FM dataflow across CEs, significantly reducing the need for off-chip FM access and achieving substantial performance improvements over previous designs. 
Nevertheless, all PEs in this architecture support only a fixed data reuse pattern, which might lead to a sub-optimal memory footprint.
Additionally, the conventional parallelism allocation method cannot coordinate all layers with a balanced workload, which inevitably impairs computational efficiency.
\parskip=0pt

The target of this paper is to reduce on-chip and off-chip memory utilization while enabling high computational efficiency for LWCNN acceleration.
To this end,
this paper proposes a streaming architecture with hybrid CEs for memory optimization and a balanced dataflow strategy for computing optimization.
Based on the aforementioned data distributions, FRCEs and WRCEs are introduced in the architecture.
The former completely removes off-chip memory access with a low cost of on-chip buffer size in shallow layers, while the latter largely decreases off-chip data traffic in deep layers.
Compared with the design in \cite{jiang2023high}, which employs only a single data reuse scheme, 
this work supports hybrid data reuse schemes and consumes only about half the BRAMs for the same model.
In addition, the balanced dataflow strategy optimizes data congestion and allocates fine-grained parallelism for the layer-specific CEs, significantly enhancing the computing efficiency of PE resources. 
The details of the proposed work are to be discussed in the following sections. 

\parskip=0pt

%


%% file: chapters/Architecture.tex

\begin{figure}[!t]
\centerline{\includegraphics[width=\linewidth]{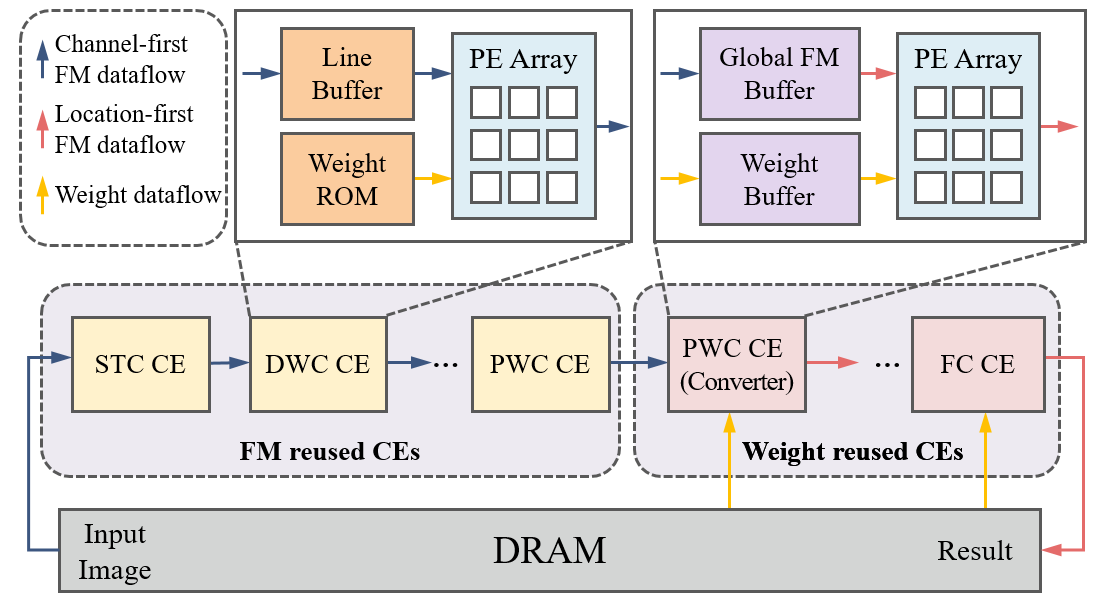}}
    \caption{Architecture of the proposed accelerator.}
    \label{fig:accelerator}
\end{figure}

\section{Accelerator with Hybrid Computing-Engines}\label{sec:acc}
\subsection{Architecture Overview}

Witnessing the significantly increased memory access demand for FMs in LWCNNs, this paper proposes a novel streaming accelerator with hybrid CEs to minimize both on-chip and off-chip memory costs. As depicted in Fig.~\ref{fig:accelerator}, each layer in the LWCNN is processed by a dedicated CE so that the results of each CE can be directly transferred to the subsequent CE instead of being transferred out of the chip. Hence, the memory access stress of low-computational density structures, e.g., DSC and SCB, can be mitigated effectively. Based on the analysis of memory distribution in Section~\ref{sec:distribution}, the proposed accelerator further classifies all CEs into two types: FRCEs for shallow layers and WRCEs for deep layers. In FRCEs, which store all weight data on-chip and thus eliminate off-chip memory access, we use a line buffer to store intermediate activations and overlap the computation between adjacent layers. A fully reused feature map scheme is adopted to further reduce line buffer size and computational delay. For WRCEs, which correspond to deep layers with a large number of parameters, a fully reused weight scheme is introduced for maximum weight reuse to minimize external memory traffic for weight data. The ping-pong strategy is applied for both FM and weight buffers to prevent pipeline blocking and hide off-chip communication latency. Corresponding to FRCEs and WRCEs, the transfer order of FMs is also tuned from channel-first to location-first through a converter CE at the group boundary to avoid inefficient memory bandwidth usage. 
Thanks to the collaboration of hybrid CEs in the streaming architecture, the proposed accelerator eliminates all DRAM access for intermediate FMs between CEs (non-shortcut) and significantly decreases off-chip weight traffic while maintaining a small on-chip buffer size.

\begin{figure}[!t]
\centerline{\includegraphics[width=0.78\linewidth]{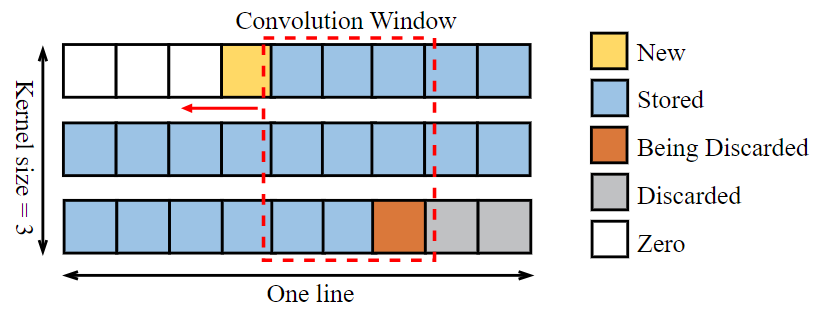}}
    \caption{Fully reused feature map scheme performed in line buffer of a $3 \times 3$ convolutional layer.}
    \label{fig:conv3}
\end{figure}

\subsection{Hybrid Computing Engines}\label{sec:hybrid ce}

Considering the significantly increased bandwidth requirements for FM in LWCNN models, traditional single-CE-based overlay architectures tend to incur high memory access overhead and potential performance limitations since activations have to be transferred to and from off-chip memory multiple times. 
Therefore, a multi-CE-based streaming architecture is introduced to alleviate this overhead by mapping each layer with a corresponding CE and transferring FMs between CEs.
However, previous streaming accelerators \cite{MALOC18, nguyen2019high, nguyen2021layer, highutilization22, jiang2023high} suffer from large on-chip memory occupation or repetitive DRAM access for weight data, which degrades the overall performance.

To decrease on-chip memory occupation and weight-induced off-chip memory traffic in conventional streaming architectures,
all layer-specific CEs are categorized into FRCEs and WRCEs, each using distinctive data reuse schemes.
The design target of FRCE is to eliminate off-chip memory access while minimizing on-chip memory usage.
Thus, the small-sized parameters in shallow layers are stored in BRAMs at a low on-chip memory cost.
Unlike weight storage with a fixed size,
the memory capacity required by intermediate FMs is determined by the lifetime of pixels being processed, which is affected by the data reuse scheme.
Note that the dataflow is transferred in a channel-first order, where a pixel is used to describe the feature data, which omits the channel dimension.
In \cite{nguyen2019high, nguyen2021layer,highutilization22},
a line-reused scheme, where the processing granularity of each CE is one line of pixels, is adopted to achieve a higher weight reuse rate.
However, each CE has to preserve an extra line of pixels to guarantee the continuity of computation.
Moreover, additional buffer space for shortcuts is also required due to the prolonged pixel life.
To further reduce the size of the line buffer, the fully reused feature map scheme is adopted in FRCE to take full advantage of the locality of the convolution operation.
Take $3 \times 3$ convolution with a stride of one as an example, as shown in Fig.~\ref{fig:conv3}.
When the first complete convolution window is cached in the line buffer, parallel computation of the adjacent layer is enabled.
The current convolution window is directly calculated across all kernels.
In this way,
the information of the oldest pixel can be fully integrated into the corresponding generated pixels in the subsequent layer, namely, its lifetime is ended and its buffer space can be overwritten by a new incoming pixel. 
As a result, for a $k \times k$ kernel size, 
the fully reused feature map scheme only needs to cache $k-1$ full lines plus $k-1$ pixels to maintain an available convolution window.
Compared to the line-based reuse scheme, this approach saves one line of buffer size even if the buffer lines increased to k full lines to reserve extra space for overlapping computations between layers.

\begin{figure}[!t]
    \centering
    \subfloat[]{\includegraphics[width=\linewidth]{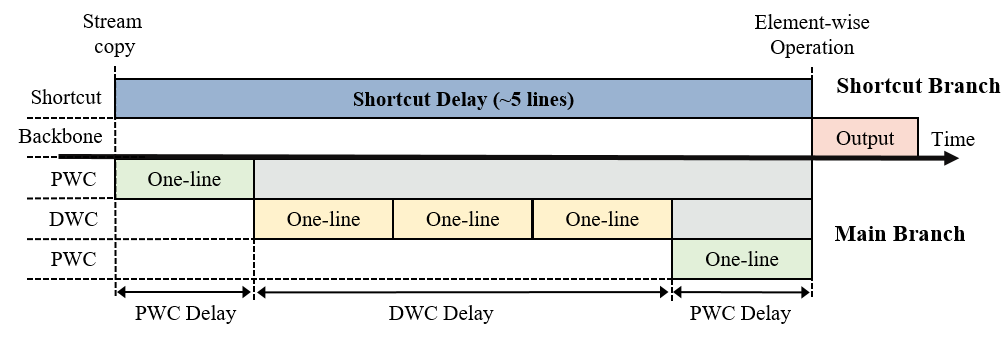}\label{fig:scblrl}}\\
    \subfloat[]{\includegraphics[width=\linewidth]{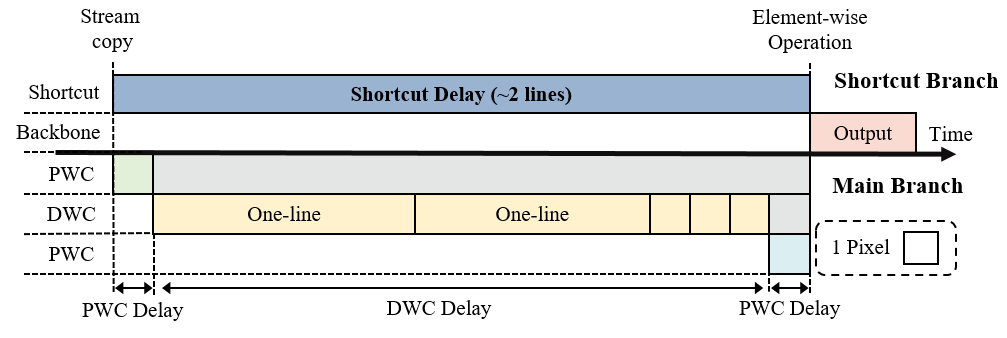}\label{fig:scbfmrl}}
     \caption{Computing timing diagram of SCB. (a) Line-based weight reuse scheme. (b) Fully reused feature map scheme.}
    \label{fig:scbrl}
\end{figure}

For shortcut branches, an extra delayed buffer is deployed to avoid additional external data transmission.
The fully reused feature map scheme not only reduces the line buffer size among CEs of the main branch but also decreases the computation delay, which further reduces the delayed buffer size in the shortcut branch.
Assuming pixels are transmitted into the accelerator continuously, and the intermediate data is uniformly transferred between CEs.
Fig.~\ref{fig:scbrl} shows the timing diagram of the implementation for the SCB in Fig.~\ref{fig:data distribution}(b) using the two methods 
mentioned above, where the upper and lower parts represent the latency of the shortcut branch and the main branch, respectively.
For the line-based weight reuse scheme, the computation can start only when $k$ full lines are buffered.
To guarantee a balanced latency between two branches and avoid system deadlock, a buffer size of at least five lines of pixels is required in the delayed buffer to achieve data synchronization, resulting in a total of thirteen buffer lines in the entire SCB structure.
By comparison, 
only about two lines of pixels need to be stored in the shortcut branch and four lines in the total SCB structure thanks to the lifetime reduction of pixels in FRCE. 
This represents a 69.23\% reduction of FM buffer size compared to the conventional method.
Consequently, the consumption of on-chip storage resources can be significantly reduced by fully reusing convolutional windows instead of the line-based reuse scheme in the existing accelerators.

For deep layers with small FMs and a large number of parameters,
traditional streaming accelerators \cite{blott2018finnr, MemoryEff20} tend to store all weights into on-chip BRAMs, 
which incurs unbearable area overheads and makes it difficult to scale well for different networks.
Recent state-of-the-art methods \cite{highutilization22, jiang2023high} store weight data of deep layers in external DRAM due to limited on-chip memory size and their low reuse rates. 
Although this approach effectively reduces on-chip memory occupation as more layers are stored in DRAM, it causes repetitive off-chip weight access due to the fixed computing pattern deployed among all CEs,
which may further impact system performance if off-chip bandwidth demand becomes a new memory bottleneck.
In our manner,
to minimize off-chip memory traffic in these layers, a fully reused weight scheme is introduced in the weight-reused CE for maximum kernel data reuse.
To this end,
a ping-pong structure-based global FM buffer is deployed in STC and PWC layers to store the FM data entirely. 
This allows each kernel load from external memory to be directly calculated across all FMs, enabling only one-time access for external parameters.
For DWC layers, which merely require calculations on a single channel dimension,
only a single channel of feature maps with a certain number of lines needs to be stored to form a complete window for computation since the FM is transferred with a location-first dimension in the weight reused scheme.  
Consequently, with the fully reused weight scheme in deep layers, redundant weight access can be eliminated at a low cost of on-chip memory.

Based on the above analysis, Table \ref{tab:ce comparison} provides a comparative summary of FRCE and WRCE.
These two groups of CEs can be allocated according to the network structure and available resources of the target platform to achieve the optimal trade-off between on-chip memory occupation and off-chip memory bandwidth.
The allocation algorithm is illustrated in Section~\ref{sec:mem alloc}.

\begin{table}[!t]
\renewcommand{\arraystretch}{1.2}
\begin{center}
\resizebox{\linewidth}{!} {%
\begin{threeparttable}
\caption{ Comparison Between Proposed CEs}
\label{tab:ce comparison}
\begin{tabular}{|c|c|c|}
\hline
Features                & FRCE               & WRCE                  \\ \hline
Reuse scheme            & Fully FM reuse     & Fully weight reuse   \\ \hline
Minimize FM buffer size & $(K-1)\times F+K-1$ & $2 \times F^2 \times M$\tnote{*} \\ \hline
Weight storage          & On-chip            & Off-chip             \\ \hline
Weight reads            & $F^2$                & 1                    \\ \hline
Shortcut implementation & Delayed buffer     & Off-chip storage     \\ \hline
Off-chip memory access  & 0                  & Weights and shortcuts  \\ \hline
Suitable layer          & Shallow layers     & Deep layers          \\ \hline
\end{tabular}%
\begin{tablenotes}    
        \footnotesize              
        \item[*] The FM buffer size is negligible for DWC layers.       
      \end{tablenotes}           
\end{threeparttable}
}
\end{center}
\end{table}
\begin{figure}[!t]
\centerline{\includegraphics[width=\linewidth]{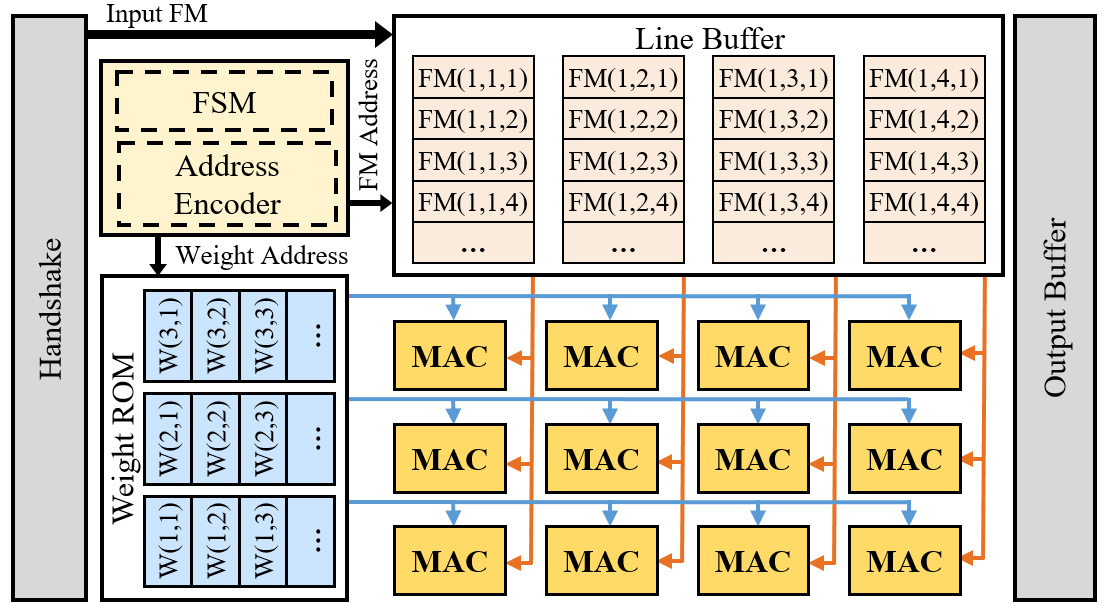}}
    \caption{Structure of FRCE. WRCE and FRCE have a similar structure except the feature map and weight buffer method.}
    \label{fig:CE structure}
\end{figure}
\subsection{Layer-Specific CE Design} 
\subsubsection{Adaptive bandwidth computing engine} 

\begin{figure}[!t]
    \centering
    \subfloat[]{\includegraphics[width=0.4\linewidth]{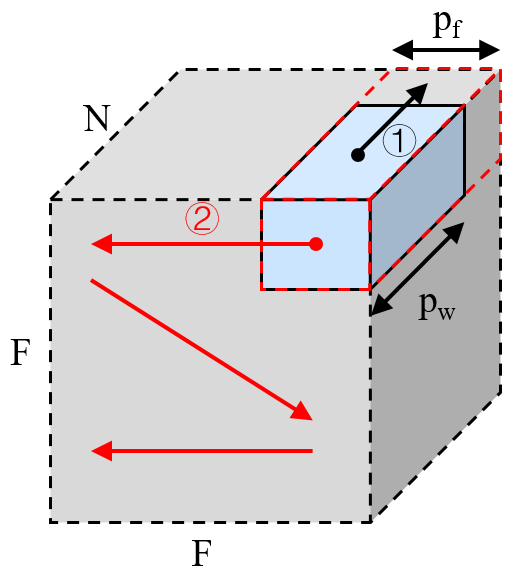}\label{fig:frce}}  ~~~~
    \subfloat[]{\includegraphics[width=0.4\linewidth]{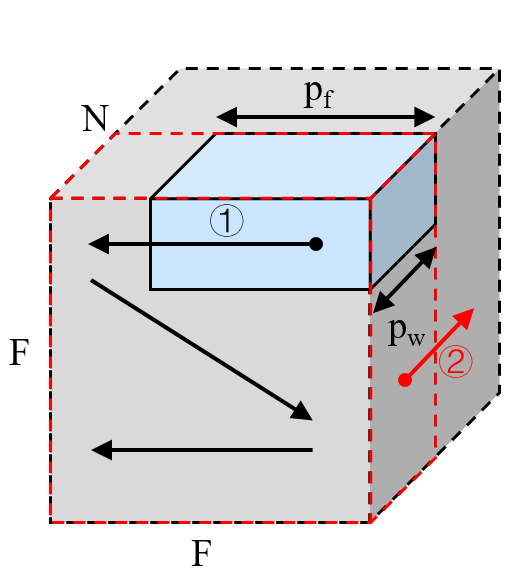}\label{fig:wrce}}
     \caption{Diagram of scheduling for output tiles. (a) FRCE. (b) WRCE.}
    \label{fig:CEout}
\end{figure}
To maintain high bandwidth for PEs while minimizing unnecessary logic resource consumption in the streaming architecture,
the layer-specific CE is designed as shown in Fig.~\ref{fig:CE structure}. 
With a handshake mechanism to synchronize computations across all layers,
the input tile transferred from the preceding CE is first tuned parallelism to match computing bandwidth and then stored in the FM buffer. 
After the convolution window is initialized,
the FSM controller schedules the computation process and passes the location information to the address encoder, which manipulates the address to prefetch data from both FM and weight buffers. 
For a given parallelism $P$ in a specific layer, 
conventional designs tend to utilize parallelism within a kernel first, which is straightforward but increases the bandwidth burden because of the inability to reuse FMs and weights.
In our accelerator, since all layers are computed in parallel, providing adequate inter-layer parallelism,
parallel computation across FM ($P_f$) and kernels ($P_w$) is employed to reduce the bandwidth requirements with the same computing resources.  
\begin{figure}[!t]
\centerline{\includegraphics[width=\linewidth]{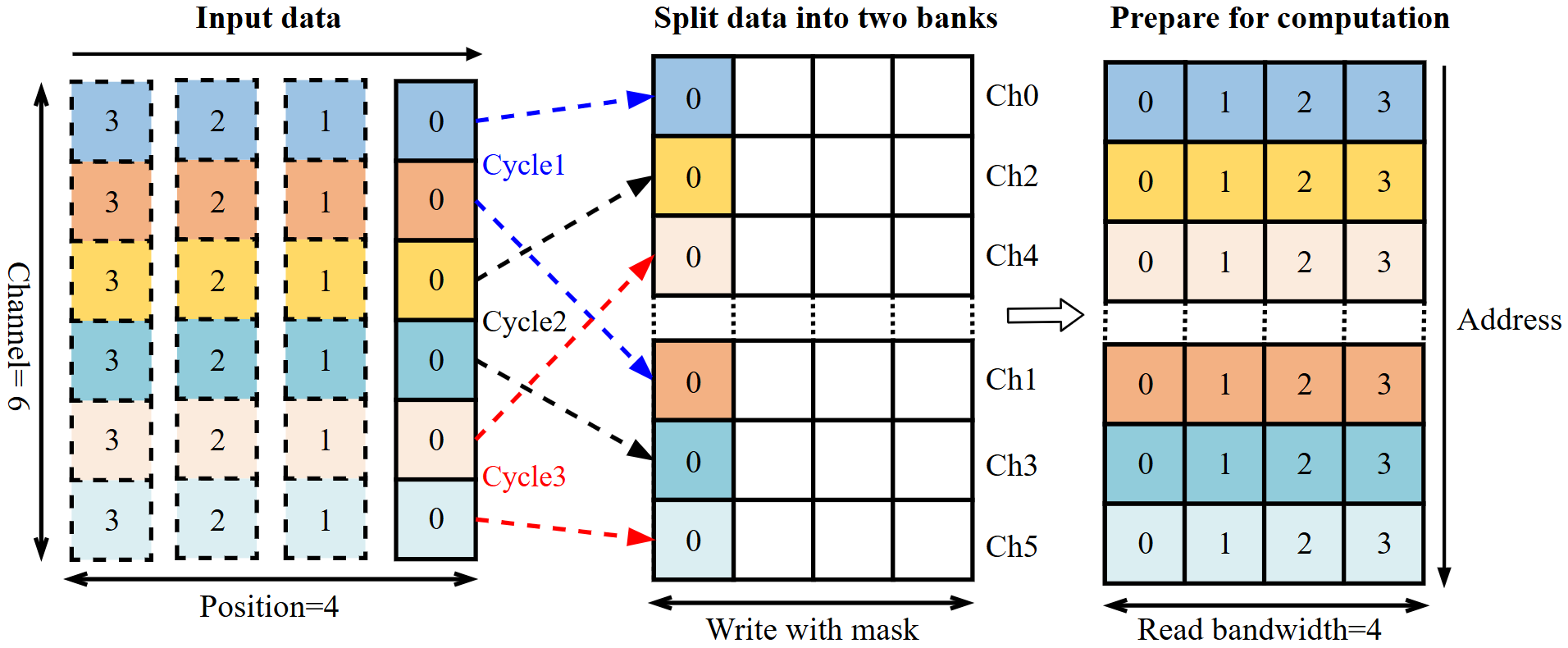}}
    \caption{Process of converting a channel-first dataflow to a location-first dataflow.}
    \label{fig:converter}
\end{figure}
During the convolution process, input FMs and weights are continuously broadcast into the PE array along the vertical and horizontal directions, respectively.
Each PE consists of a multiplier followed by an accumulator, where the accumulator aggregates the partial sums of the entire kernel and outputs the final results.
As for DWC layers, the absence of accumulation across the channel dimension allows output results to be directly obtained after sliding along the kernel map rather than the cubic kernel.
By this means,
an output tile including $P_f \times P_w$ final activations is generated after one iteration and stored in the output buffer subsequently.
As mentioned before, different reuse schemes generate results with different extended dimensions. 
To minimize output buffer size, FRCE generally prefers parallelism in the output channel dimension ($P_w$) to acquire more output channels in a single iteration. Conversely, WRCE prefers parallelism in the input FM dimension ($P_f$) to get a larger scope of output FMs.
After the current FM windows are completely processed, namely, $P_f \times N$ activations are buffered and reshaped, these values in the output buffer are sent to the next CE for consecutive computation. 
In most layers, only parallelism in the output channel dimension is utilized, which means $P_f = 1$, allowing computation results to be directly transferred to the next layer.
Hence, the output buffer can be avoided and the on-chip storage resource consumption is further reduced. 
These processes are repeated to calculate the subsequent $P_f$ convolution windows until the entire output FM is generated.
In terms of WRCE, a similar computation procedure with reversed extended dimensions is shown in Fig.~\ref{fig:CEout}\subref{fig:wrce}.

\subsubsection{Dataflow order converter}

Due to the different order of FM dataflow between FRCE and WRCE, the first WRCE at the group boundary needs to support the channel-first dataflow from FRCE.
However, a mismatch of dataflow order may lead to insufficient bandwidth and complex control logic.
To maintain order consistency in the subsequent FM dataflow, 
an order conversion method is introduced, as shown in Fig.~\ref{fig:converter}.
Multiple banks of RAM with write masks are employed to decompose and rearrange the input data.
The input data is first serialized and then written to each bank with corresponding masks.
FM data from the same channel are distributed across different banks or addresses, 
while data from the same location slice are written to the same address so that they can be fetched in the same cycle.
By this means, data order transpose is achieved without additional storage space and data management is greatly simplified.

%% file: chapters/Dataflow.tex
\section{Accelerator with Balanced Dataflow}
Based on the proposed memory-optimized architecture in Section~\ref{sec:acc},
a balanced dataflow strategy is specially designed to improve the computing efficiency of our streaming accelerator.
It can be divided into two levels:
1) The fine-grained parallel mechanism to enable large parallel space and efficient computing resource allocation at the theoretical level.
and 
2) a dataflow-oriented line buffer scheme to mitigate the on-chip bandwidth supply problem caused by padding data and large convolutional stride at the implementation level.

\subsection{Fine-grained Parallel Mechanism}\label{sec:fine}    
\begin{figure*}[!t]
\centerline{\includegraphics[width=\linewidth]{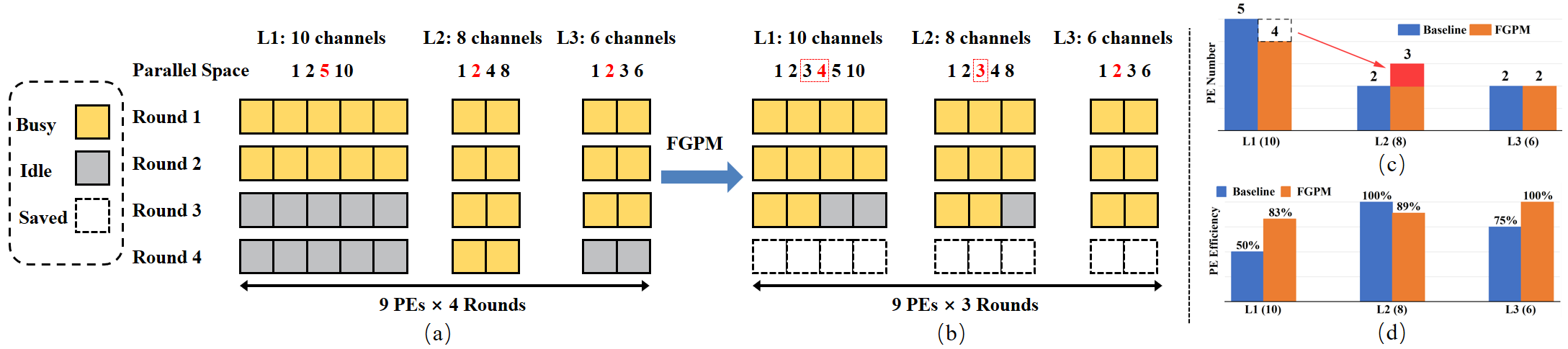}}
    \caption{Parallelism allocation and comparisons in different granularity:  (a) Parallelism allocation with factorized granularity with 9 parallel PEs; (b) Parallelism allocation with the proposed FGPM with 9 parallel PEs; (c) PE number comparison; (d) PE efficiency comparison.}
    \label{fig:paral}
\end{figure*}

According to the barrel effect of streaming architecture, the overall throughput of our accelerator is fundamentally constrained by the computing time of bottleneck CE.
For a given network and limited parallelism (PEs), 
ideally, these computing resources should be allocated proportionally to each CE based on the required MAC operations, ensuring balanced dataflow across CEs to achieve maximum throughput. 
To this end, most of the previous streaming accelerators adopted factorized granularity to determine the parallelism of CEs, namely, choosing parallelism from the factors of dimensional maximum parallelism.
Although this method reduces design complexity, it offers only a small number of integer factors, resulting in unbalanced dataflow and inefficient utilization of PEs.
A simplified example with three layers and a single parallel dimension (output channel) performed by 9 PEs is depicted in Fig.~\ref{fig:paral}(a),
where the parallel space represents the number of available parallelism configurations that can provide different computing times.
Note that the current allocation achieves the highest throughput in conventional factorized granularity.
For non-bottleneck layers in streaming architecture, i.e., L1 and L3,
the parallelism requirement driven by bottleneck computing time can normally be an arbitrary value due to diverse cross-layer computational requirements in LWCNN.
Restricted by factorized granularity, these non-bottleneck layers can only choose parallelism that exceeds the bottleneck time requirement from a limited range.
However, these excess parallelisms are wasted since they make no contribution to system throughput.
On the other hand, even if the bottleneck layer L2 can achieve a theoretical resource utilization of 100\% at the current parallelism, 
enhancing its throughput requires significant additional PE resources due to the large parallelism gap required for shorter computing times. 
For example, the system throughput cannot be improved even if an additional PE exists in Fig.~\ref{fig:paral}(a),
which makes additional computing resources remain underutilized until a resource increment threshold is reached.
This results in a heavy staircase effect between the number of PEs and the corresponding throughput.

To address this problem, a fine-grained parallelism mechanism is proposed to extend the parallel space in each parallel dimension fully.
Different from the factorized granularity, which focuses on the optimal parallelism allocation for individual layers locally, 
FGPM selects all integer values that can provide different computational times from each dimension as the available parallelism to achieve optimal overall resource allocation.
Given a parallel dimension with maximum parallelism $M$ and integer parallelism $P$, the computing rounds $T$ are calculated as:
\begin{align}
    T  &= ceil(\frac{M}{P})  
\end{align}
where the $ceil$ function ensures all required operations are processed. 
With all the different computing time $T$,
the valid range of P, i.e., the size of the parallel space, is $2 \times floor( \sqrt{M} )$, where $\sqrt{M}$ comes from all values of P when P is no more than $\sqrt{M}$ since each P results in a unique T, while the other $\sqrt{M}$ derives from the corresponding division mapping of the first group.
Since the factor number of M cannot exceed $2 \times floor( \sqrt{M} )$, the FGPM approach definitely offers a higher or at least equivalent parallel space compared to factorization methods, especially for M with large values or sparse factors.
For instance, using common output channel numbers like 32, 64, 128, 256, and 512, the size of parallel space can be increased by 67\%,114\%,175\%,244\%, and 340\%, respectively.
With a low hardware cost, the non-factor parallelisms are implemented by dimension padding.
The padded maximum parallelism maintains the regularity of hardware computation, and excess intermediate results are reorganized and discarded when transferred to the next CE.
In this way,
FGPM provides more available parallelism for efficient allocation of computing resources, which helps to find the optimal resource allocation scheme close to the ideal situation.

As a comparison to the previous example, Fig.~\ref{fig:paral}(b) shows the additional parallelism not only conserves PEs for non-bottleneck layers by reducing over-allocation but also alleviates the staircase effect in bottleneck layers by reducing the parallelism gap. 
In this way,
the saved PEs from L1 can be reallocated to the slowest layer, L2, to achieve a relatively uniform and balanced dataflow.
Fig.~\ref{fig:paral}(c) and Fig.~\ref{fig:paral}(d) further illustrate the comparison of PE numbers and efficiency between these two methods.
Despite a slight efficiency reduction at the bottleneck layer L2, the efficiency of the non-bottleneck layers is greatly improved thanks to the fine-grained resource allocation strategy, which effectively promotes the overall throughput with the same hardware resources.

\subsection{Dataflow-oriented Line Buffer Scheme}  

\begin{figure}[!t]
\centerline{\includegraphics[width=\linewidth]{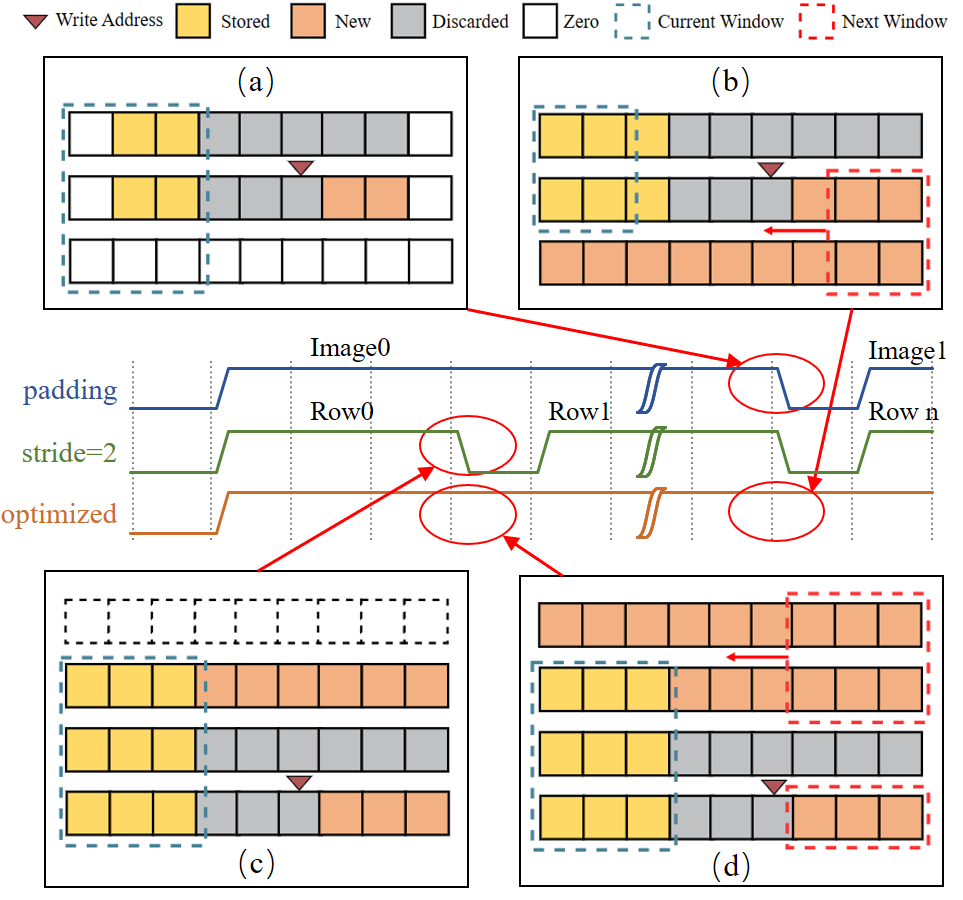}}
    \caption{Timing diagram of the PE working state and the corresponding FM organization in the line buffer:  
    (a) Padding data is written directly into the line buffer for processing, causing an interruption in bandwidth supply;
    (b) No padding data is written to the line buffer; the padding operation is completed when the data is sent to PE;
    (c) Processing $k \times k$ convolutional layer with a large stride directly in a line buffer with a $k$ row size, causing a bandwidth supply interruption;
    (d) Processing $k \times k$ convolutional layer with a large stride in an optimized line buffer with extra buffer size; 
    }
    \vspace{-0.2cm}
    \label{fig:bufferopt}
\end{figure}

Another crucial issue affecting the computational efficiency of streaming architectures is data congestion, which is mainly caused by inefficient padding implementation and a mismatch between buffer organization and convolutional stride. 
Existing works \cite{RealTimeObject18, yan2021fpga, LowPowerMobil23} normally insert zeros directly into the input FM dataflow before feeding them into the line buffer, treating padding data as general pixels during convolution operations.
This comes with two issues: data backpressure for the preceding CE and invalid convolution windows in the current CE. 
First, writing padding values to the line buffer consumes its write bandwidth, resulting in either a large input buffer to store input pixels temporarily or a computing stall in the preceding CE. Additionally, padding values, which have low information density, occupy precious memory space intended for input pixels, leading to invalid convolution windows during the image-switching process. Fig.~\ref{fig:bufferopt}(a) illustrates the direct insertion approach in a $3 \times 3$ convolution layer with padding on all sides, where padding data is reused in both the last row of the previous image and the first row of the next. 
After generating the last window in the previous image, the new FM window is delayed due to data congestion. 
This necessitates buffering an additional line of input pixels in the line buffer to form the first active window in the following image, causing interrupted data supply.
A similar issue arises when processing layers with large strides. As shown in Fig.~\ref{fig:bufferopt}(c), a large stride (e.g., stride of 2) prevents the timely generation of new convolution windows after completing computation on the previous line, which inevitably leads to idle states of PEs. 

To address the bandwidth supply problem caused by data congestion, a dataflow-oriented line buffer scheme is proposed to enhance hardware computing efficiency by ensuring sufficient bandwidth supply. 
Considering the uniformity of padding data and its specific insertion locations,
the padding operation is performed by the address controller within the CE while the pixels are sent to the PE array instead of being written into the line buffer as input pixels. This approach prevents congestion in upstream data.
Additionally,    
padding data generated by the address logic increases input bandwidth indirectly, eliminating window bubbles during image switching. 
As shown in Fig.~\ref{fig:bufferopt}(b), since the line buffer is not occupied by padding data, a new line of pixels has been stored in the line buffer when the previous image is completed. 
Combined with the supplemented padding data, a new convolutional computation round for the subsequent image can start up immediately, significantly reducing idle hardware resources. 
For convolution layers with large strides, as shown in Fig.~\ref{fig:bufferopt}(d), an additional line buffer is allocated in our scheme to eliminate window bubbles caused by the mismatch between limited buffer space and required data scope. With the optimized padding generator and additional buffer space, the computing efficiency of STC and DWC layers with diverse padding locations and strides can be effectively improved.

%% file: chapters/Allocation.tex
\section{Resource-Aware \\ Memory and Parallelism Allocation}

To further improve the performance and compatibility of our streaming accelerator,
this section proposes a resource allocation methodology based on the quantitative analysis of the throughput and memory access in the presented architecture. 
It can be decomposed into two parts:
1) the balanced memory allocation algorithm to identify the group boundary to minimize off-chip memory (DRAM) access while meeting on-chip memory (SRAM) constraints; 
2) the dynamic parallelism tuning algorithm to distribute PEs for each layer to achieve maximum throughput with a certain number of DSPs.

\subsection{Balanced Memory Allocation} \label{sec:mem alloc}

The memory utilization of the accelerator depends on the number of FRCEs and WRCEs, which is directed by the location of the group boundary. 
A deeper location increases the number of FRCEs, reducing off-chip access but may lead to excessive on-chip storage requirements. 
In contrast, despite a higher number of WRCEs removing more on-chip weights, off-chip access and computing latency are accordingly increased, which also degrades system efficiency. 
Therefore, it is crucial to find the optimal trade-off between these two types of CEs based on the performance model and design constraints.

Assuming the given network has L layers with l layers deployed as FRCEs.
According to the analysis in Section~\ref{sec:hybrid ce},
the total SRAM size can be roughly calculated as:
\begin{align}
  SRAM_{total} &= SRAM_{FMRL} + SRAM_{WRCE}+ SRAM_{SCB}  \notag\\
   &= {\textstyle \sum_{i=0}^{l}(Line\_buffer(i)  + Weight\_ROM(i))}  \notag \\
    &+  {\textstyle \sum_{i=l}^{L}(GFM\_buffer(i)  + Weight\_buffer(i))} \notag \\
    &+{\textstyle \sum_{i=0}^{l}Shortcut\_buffer(i)} 
\end{align}
where $Line\_buffer_{i}$, $GFM\_buffer_{i}$, and $Weight\_ROM_{i}$ generally occupy the main storage space in FRCEs and WRCEs.
It's worth noting that the line buffer is not required in PWC layers since they do not involve inter-pixel correlation operations.
Similarly, since there is no cross-channel operation in DWC layers, the global buffer only needs to store the FMs of a single channel, which favors both SRAM size and layer latency. 
The sizes of the weight buffer ($Weight\_buffer(i)$) and the shortcut buffer ($Shortcut\_buffer(i)$) are relatively small. The former depends on the weight parallelism, while the latter is directly influenced by computing latency and network structure.
Without considering the input image and final results, the DRAM access per image in our accelerator can be calculated as follows:
\begin{align}
  DRAM_{total}  &= DRAM_{WRCE} + DRAM_{SCB}  \notag\\
                &= \textstyle \sum_{i=l}^{L}({Weight(i)+Shortcut(i)}) 
\end{align}
where $Weight(i)$ represents the weight size in non-DWC layers, and $Shortcut(i)$ is twice the size of the FMs in an SCB structure.
Note that there is no off-chip access for intermediate activation between CEs thanks to the streaming architecture.
\begin{algorithm}[t] \footnotesize
	\caption{Balanced Memory Allocation Algorithm}
        \label{algo:layer allocation}
        \SetKwInOut{Input}{Input}
  \SetKwInOut{Output}{Output}
            \Input{$SRAM_{FPGA}$  \quad /*Available SRAM size.      */}
            \Output{$NUM_{FRCE}$  ~\quad   /*Maximum number of FRCE.       */}
            
            Initialization:$NUM_{FRCE}=0$
    
    /*First iteration  ~~~\quad\quad     */
    
    \For{Layer $i = 1:L$}{   
     
     Calculate $SRAM_{FRCE}(i), SRAM_{WRCE}(i)$
     
       \If{$SRAM_{FRCE}(i) <= SRAM_{WRCE}(i)$}{
            
            Layer(i) $\Rightarrow $ FRCE 
            
           $NUM_{FRCE} = i$ }
          
       \Else{
          \textbf{break}
          } 
    }
    
     Export configuration with minimum SRAM size
     
     /*Second iteration  \quad\quad         */
     
        \For{Layer $ i = (NUM_{FRCE}+1):L$}{
           Layer(i) $\Rightarrow $ FRCE
           
          Calculate $SRAM_{total}$ 
          
            \If{$SRAM_{total} < SRAM_{FPGA}$}{
                  $NUM_{FRCE} = i$ 
            }
            \Else{
             
             \textbf{break}  
             
            }
        
        }  
    
    Update configuration 
  \end{algorithm}

According to the memory evaluation model, Algorithm~\ref{algo:layer allocation} describes the procedure with two rounds of iteration to determine the number of FRCEs, namely, the location of the group boundary. The first iteration aims to find the scheme that uses the smallest SRAM size. 
This is achieved by incrementally increasing the number of FRCEs layer by layer until reaching a specific layer where the SRAM consumption with FRCEs exceeds that with WRCEs. 
Given the typical distribution of FM and weight sizes, this configuration is considered to represent the minimum requirement of SRAM size. 
For modern FPGAs with fixed and rich memory resources, if there is still available SRAM on the target chip, the second iteration will continue to advance the boundary layer based on the results of the first iteration until the utilized SRAM size exceeds the threshold. Note that the SRAM footprint is only an approximate estimate based on the BRAM number. 
In this way, our accelerator can fully utilize available on-chip memory to minimize off-chip access and computing latency, accommodating various deployment scenarios across different FPGA chips and target networks. 
In extreme scenarios with abundant memory resources or for small networks, the entire model can be deployed with FRCEs, enabling all weights and intermediate FMs to be stored on-chip, thereby eliminating the demand for external bandwidth during computation.

\subsection{Dynamic Parallelism Tuning} 
Once the boundary layer is established, the next challenge is efficiently allocating parallelism to maximize overall throughput with given computing resources. The theoretical throughput (GOPS) of the proposed accelerator can be expressed as follows:
\begin{align}\label{equ:computetime}
 Throughput         = \frac{O_{total} \times 2}{\max_{i=1}^L{T(i)}} = \frac{O_{total} \times 2}{\max_{i=1}^L({\frac{O(i)}{P_w(i) \times P_f(i)}})}
\end{align}
where $O_{total}$ is the number of MACs of the target model, $T(i)$ and $P(i)$ represent the computing time and parallelism of layer $i$, respectively.
Note that $O(i)$ denotes the number of MACs in layer $i$ after dimension padding caused by the FGPM in Section~\ref{sec:fine}.
To equalize the theoretical throughput of each layer,
some schemes calculate the required parallelism directly based on the computational density of each layer and then fine-tune it manually.
However, limited parallel space makes it difficult to match the required parallelism accurately, which may result in sub-optimal resource allocation. Additionally, the proposed FGPM and DSP decomposition further complicate the implementation process, reducing the feasibility of accommodating various platforms and networks.

\begin{algorithm}[t]\footnotesize
	 \SetKwInOut{Input}{Input}
  \SetKwInOut{Output}{Output}
	\caption{Dynamic Parallelism Tuning Algorithm}
	\label{algo:dsp allocation}
            \Input{$DSP_{FPGA}$  \qquad \qquad /*Available DSP resource. */}
            \Output{$P_w(i), P_f(i)$  ~~\quad\qquad /*Parallelism for each layer. */}
           
            Initialization:$DSP_{total}=0$,  $P_w(i)=P_f(i)=1$, $T(i)=O(i)$
              
              \While{$DSP_{total} < DSP_{FPGA}$}{
             $T_{max} = max(T(i))$ \quad\quad /*Find bottleneck. */
             
                  \For{Layer $i = 1:L$}{
                    
                    Export $P_w(i), P_f(i)$
                    
                      \If{$T(i) == T_{max}$}{
                       
                       Increase $P_w(i), P_f(i)$ to the next level 
                       
                       Update $T(i)$
                      }
                  }
                 
                 Calculate $DSP_{total}$
            }  
\end{algorithm}

In this section,
a dynamic parallelism tuning algorithm is introduced to realize efficient computing resource allocation by incrementally adjusting DSPs for bottleneck layers.
As shown in Algorithm~\ref{algo:dsp allocation},
the parallelism of each layer is initially configured to one, making the initial computing time equal to the number of operations.
The algorithm then searches for all bottleneck layers and increases fine-grained parallelism to balance the computing time among all CEs.
Note that the increments in $P_w(i)$ and $ P_f(i)$ have different priorities depending on the CE type.
After each iteration,
the computing time of the bottleneck layers is decreased.
The iteration continues until the number of DSPs reaches the resource constraint.
Consequently,
guided by the algorithm, the accelerator achieves the highest throughput, and DSP utilization on target platforms is maximized.

Thanks to the proposed multi-CE architecture, the execution of each CE is regular and independent. Thus, the overall execution performance and resource consumption can be estimated quantitatively.
With the performance model, the resource-aware memory and parallelism allocation algorithm can explore the huge design space and enable high scalability.
As a result, the optimal implementation configuration for the given LWCNN model and FPGA chip using the proposed architecture can be obtained to maximize system performance.

%% file: chapters/Experiments.tex
\section{Experiments}
%
In this section, we first describe the evaluation results of memory occupation and computing efficiency for four mainstream LWCNNs based on the performance model. 
Subsequently, MobileNetV2 and ShuffleNetV2 are implemented with Xilinx ZC706 board as the target platform to further validate the proposed architecture.
Finally, the proposed accelerator is compared with state-of-the-art LWCNN accelerators.

\begin{figure*}[!t]
    \centering
    \subfloat[]{\includegraphics[width=0.25\linewidth]{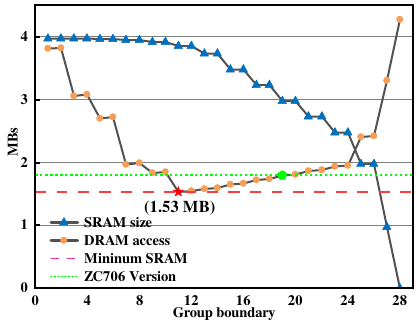}\label{fig:blaeva_sf2}}
    \subfloat[]{\includegraphics[width=0.25\linewidth]{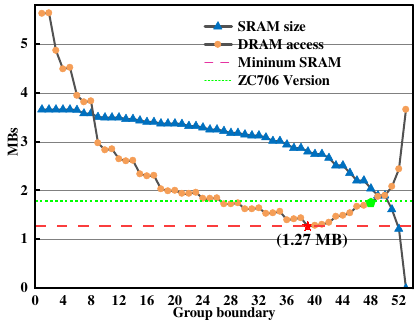}\label{fig:blaeva_sf2}}
    \subfloat[]{\includegraphics[width=0.25\linewidth]{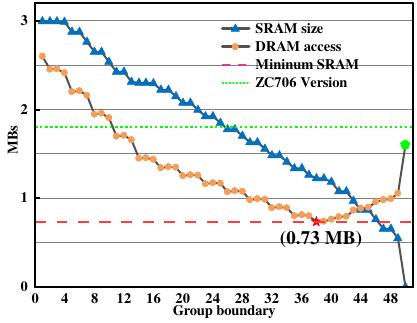}\label{fig:blaeva_sf2}}
    \subfloat[]{\includegraphics[width=0.25\linewidth]{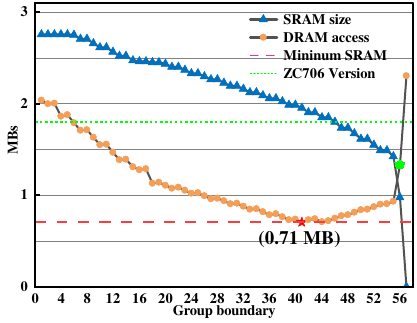}\label{fig:blaeva_sf2}}
    \caption{On-chip memory size (MB) and Off-chip memory access (MB/frame) with respect to group boundary. (a) MobileNetV1; (b) MobileNetV2; (c) ShuffleNetV1; (d) ShuffleNetV2.}
    \label{fig:mem eva}
\end{figure*}
\subsection{Experimental Setup}
To evaluate the effectiveness and generality of the proposed architecture,
four representative LWCNNs, which extensively stack DSCs and SCBs, are used as benchmarks for memory and computing simulation. 
These LWCNNs include MobileNetV1\cite{howard2017mobilenets}, MobileNetV2\cite{sandler2018mobilenetv2}, ShuffleNetV1\cite{zhang2018shufflenet} and ShuffleNetV2\cite{ma2018shufflenet}.
The input image size for all these networks is 224 $\times$ 224.
With less than 1\% loss of accuracy, both weights and activations are quantized to 8-bit precision based on methodologies from \cite{nagel2019data} and \cite{wei2023qdrop}.
ShuffleNetV2 and MobileNetV2 are implemented on the Xilinx ZC706 board to further validate the proposed architecture.
For the design constraints,
the maximum SRAM utilization and DSP utilization are empirically set at 75\% (1.80MB calculated by 545 BRAMs) and 95\% (855 DSPs), respectively, to achieve the best trade-off between timing performance and resource utilization.
The DSP decomposition is also applied in the accelerator by performing two 8 $\times$ 8 multipliers in one DSP48E1 to increase peak performance, except in the DWC layer with independent channels.
The accelerator is designed with SystemVerilog, synthesized, placed, and routed with Vivado 2021.2.

\subsection{Simulation and Implementation Results}
Based on the memory allocation algorithm described in Section~\ref{sec:mem alloc}.
Fig.~\ref{fig:mem eva} illustrates the on-chip memory size and off-chip memory access of the hybrid CE architecture with respect to the boundary location for the evaluated LWCNNs.
The balanced memory allocation algorithm sequentially searches through all layers to find the configuration that minimizes DRAM access while satisfying the SRAM constraints.
It can be observed that the SRAM size follows a U-shaped pattern as the group boundary advances, while DRAM accesses gradually decrease.
This phenomenon occurs because deploying shallow layers with FRCE favors both SRAM and DRAM utilization due to the extensive reuse of intermediate activation.
However, as the group boundary reaches the deeper layers with larger weights, 
the benefit provided by the line buffer is insufficient to counteract the negative impact of the weight ROM, which aggravates the SRAM consumption.
Correspondingly, the SRAM size reaches its minimum at this configuration.
For the target FPGA, the group boundary can be further advanced until reaching the SRAM constraints to further reduce off-chip access, as shown in the ZC706 version.
In subsequent discussions, the accelerator with the minimum SRAM size will be used as the default configuration for performance comparison.

\begin{figure}[!t]
\centerline{\includegraphics[width=\linewidth]{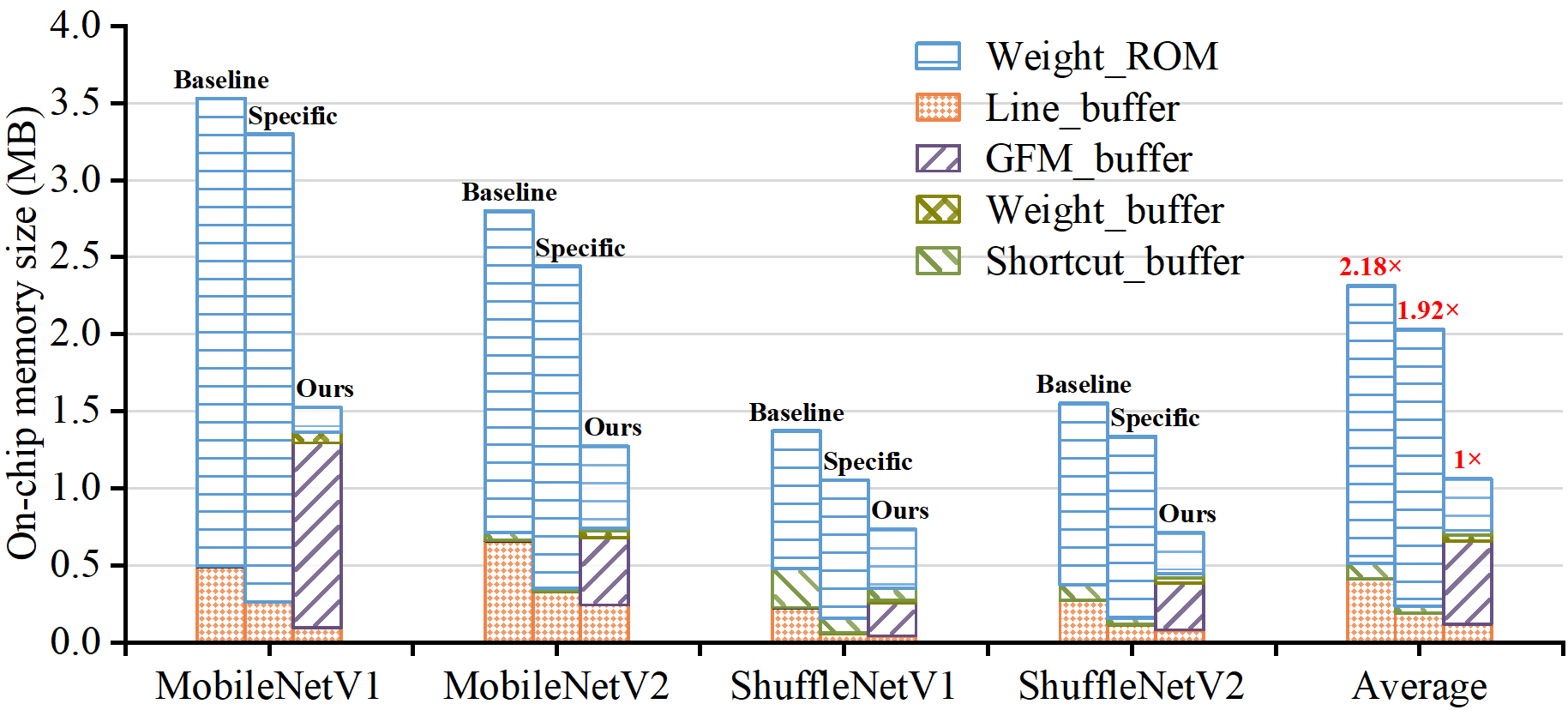}}
    \caption{On-chip memory size comparison among different streaming architectures.}
    \label{fig:sram cmp}
\end{figure}

Fig.~\ref{fig:sram cmp} presents the on-chip memory size comparison of the proposed design over other streaming architectures, where ``baseline'' means processing LWCNNs with the line-based weight reuse strategy, while ``specific'' represents using a fixed fully reused FM strategy for all convolution layers.
Note that the weights of FC layers are excluded from the evaluated schemes to ensure a reasonable comparison.
The results show that the size requirements for the line buffer and SCB buffer in the ``specific'' scheme are reduced by an average of 53.71\% and 60.0\%, respectively,  compared with the ``baseline'' scheme.
This reduction demonstrates that a shorter data life leads to a positive effect on the FM buffer size in streaming accelerators.
However, both schemes incur substantial SRAM consumption for weight data due to their fixed data reuse strategies. 
In contrast, the proposed architecture, which employs a hybrid data reuse scheme, achieves an average weight storage reduction of 81.37\% compared to the other schemes.
Even though the fully reused weight scheme requires additional GFM buffers, 
the overall SRAM footprint is still significantly reduced compared to existing schemes.

\begin{figure}[!t]
\centerline{\includegraphics[width=\linewidth]{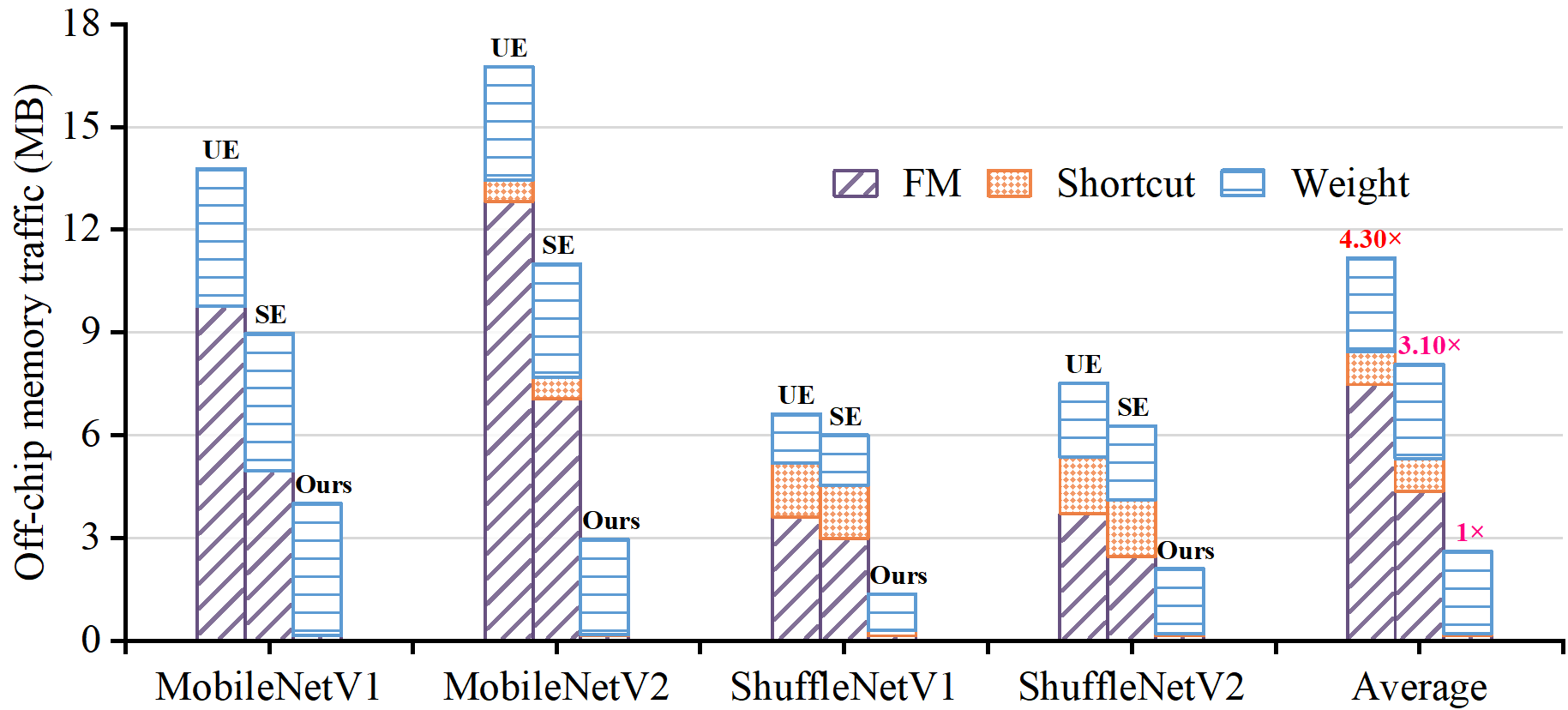}}
    \caption{Off-chip memory access comparison among UE, SE and proposed architecture.}
    \label{fig:dram cmp}
\end{figure}

Fig.~\ref{fig:dram cmp} exhibits the off-chip memory traffic comparison among unified CE (UE), separated CE (SE), and the proposed architecture during each inference of the LWCNN process,
where shortcut data is presented separately from FM data for better illustration.
Without loss of validity, we assume that all data in the UE architecture are accessed off-chip exactly once, while the SE architecture further eliminates the FM access for DWC layers. 
For LWCNNs with heavy FM access requirements, the proposed streaming architecture further extends the SE architecture so that all intermediate FM data can be transferred among CEs, resulting in an average FM access reduction of 98.07\% and 96.69\% compared to the UE and SE architecture respectively.
Additionally, the access requirements for shortcut and weight data are also reduced by 93.30 \% and 12.56\% on average, respectively, thanks to our hybrid data reuse strategy.

\begin{figure*}[!t]
    \centering
    \subfloat[]{\includegraphics[width=0.25\linewidth]{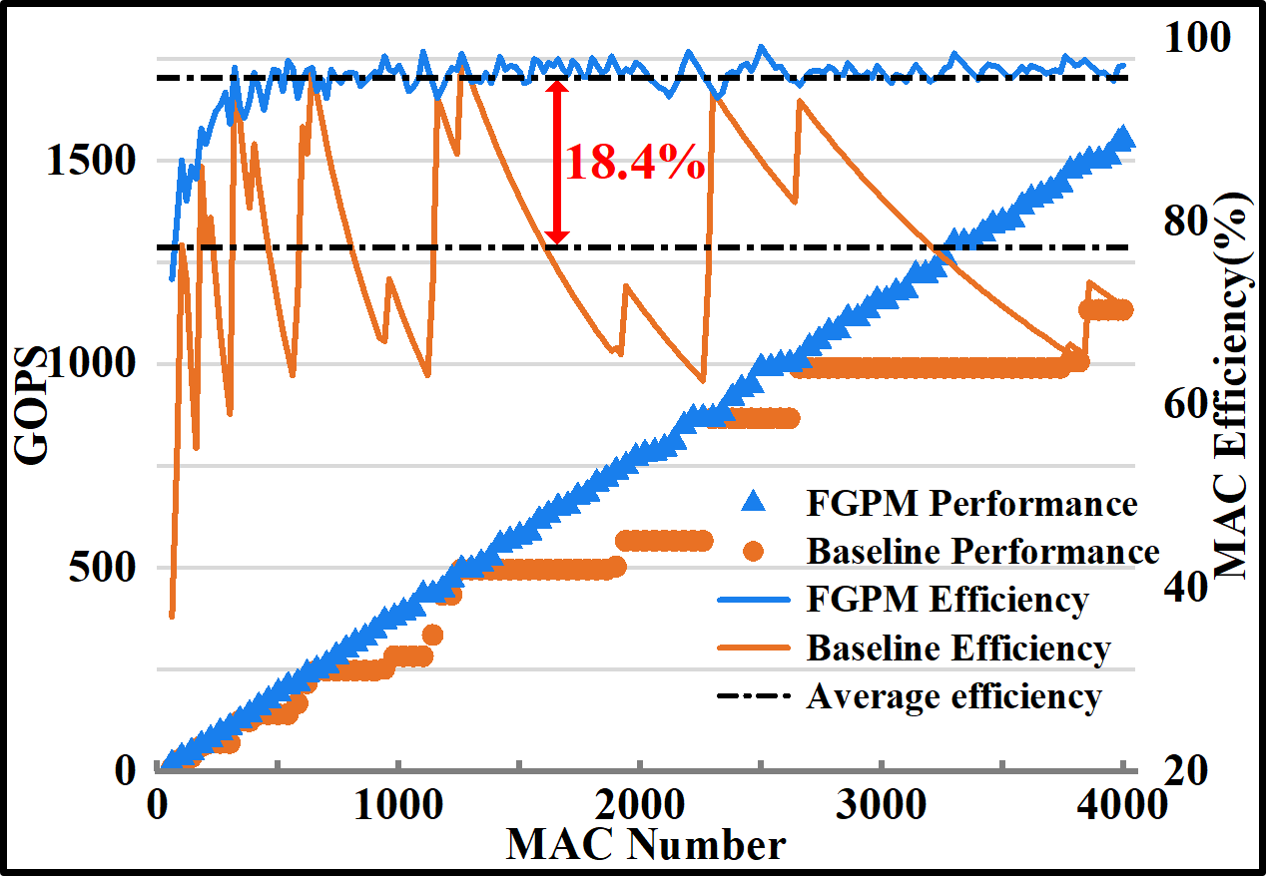}} ~
    \subfloat[]{\includegraphics[width=0.25\linewidth]{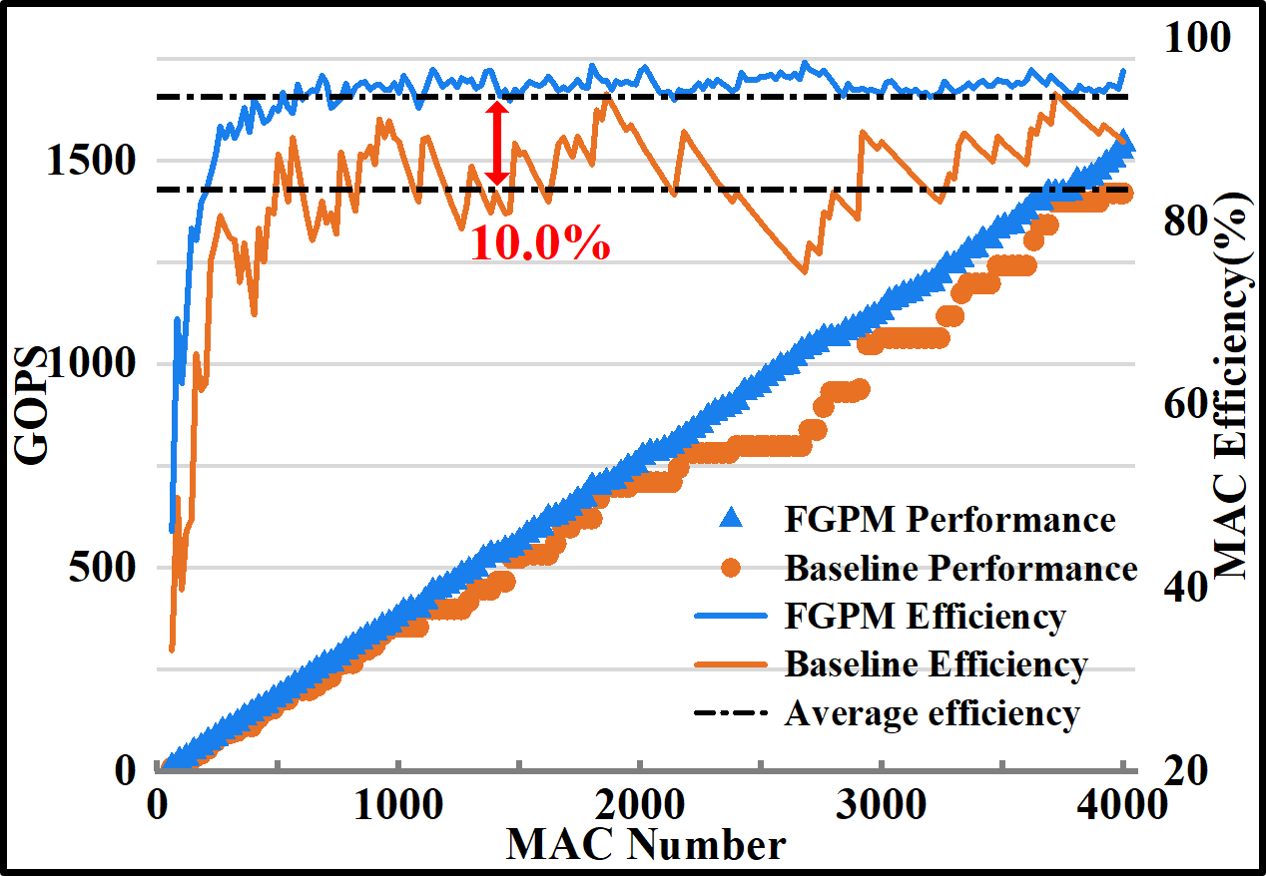}} ~
    \subfloat[]{\includegraphics[width=0.25\linewidth]{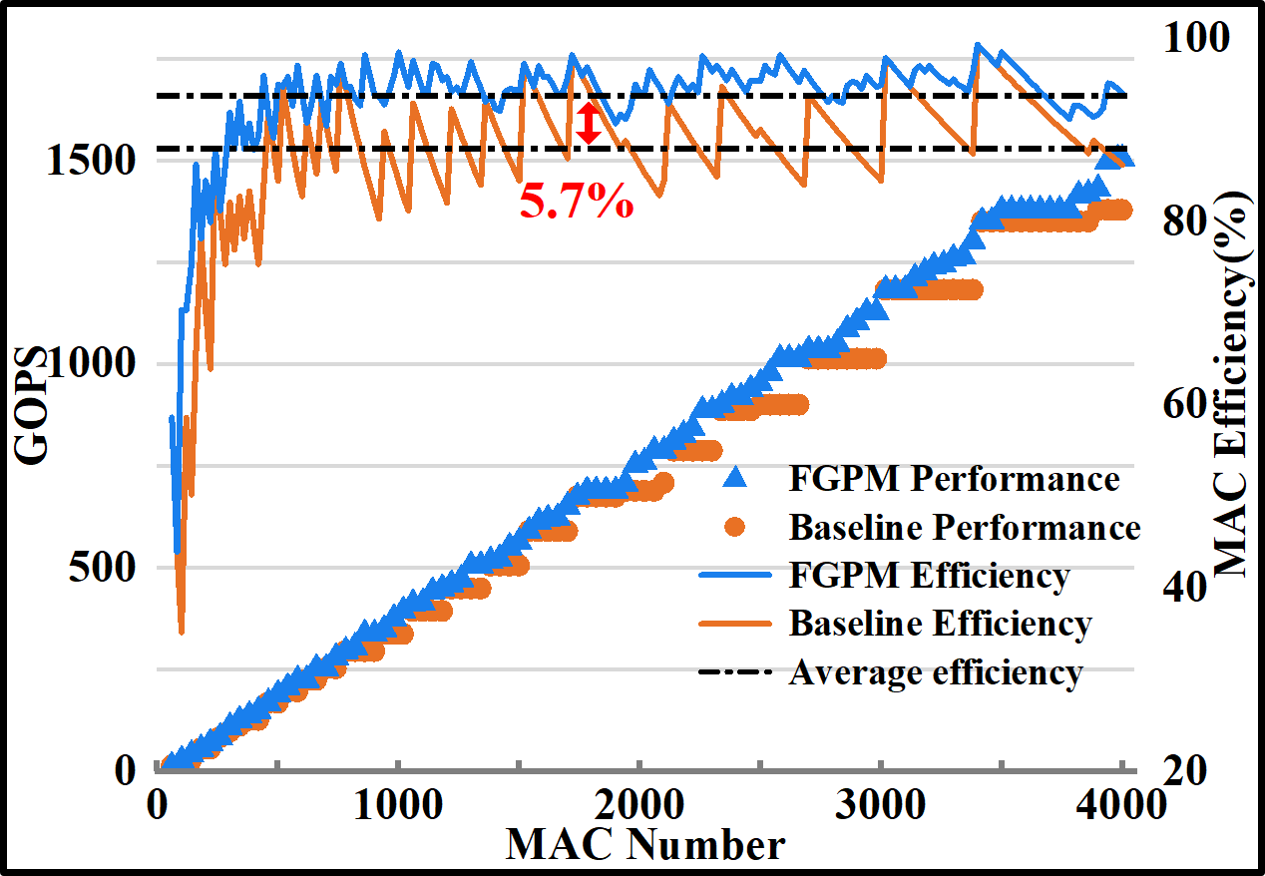}} ~
    \subfloat[]{\includegraphics[width=0.25\linewidth]{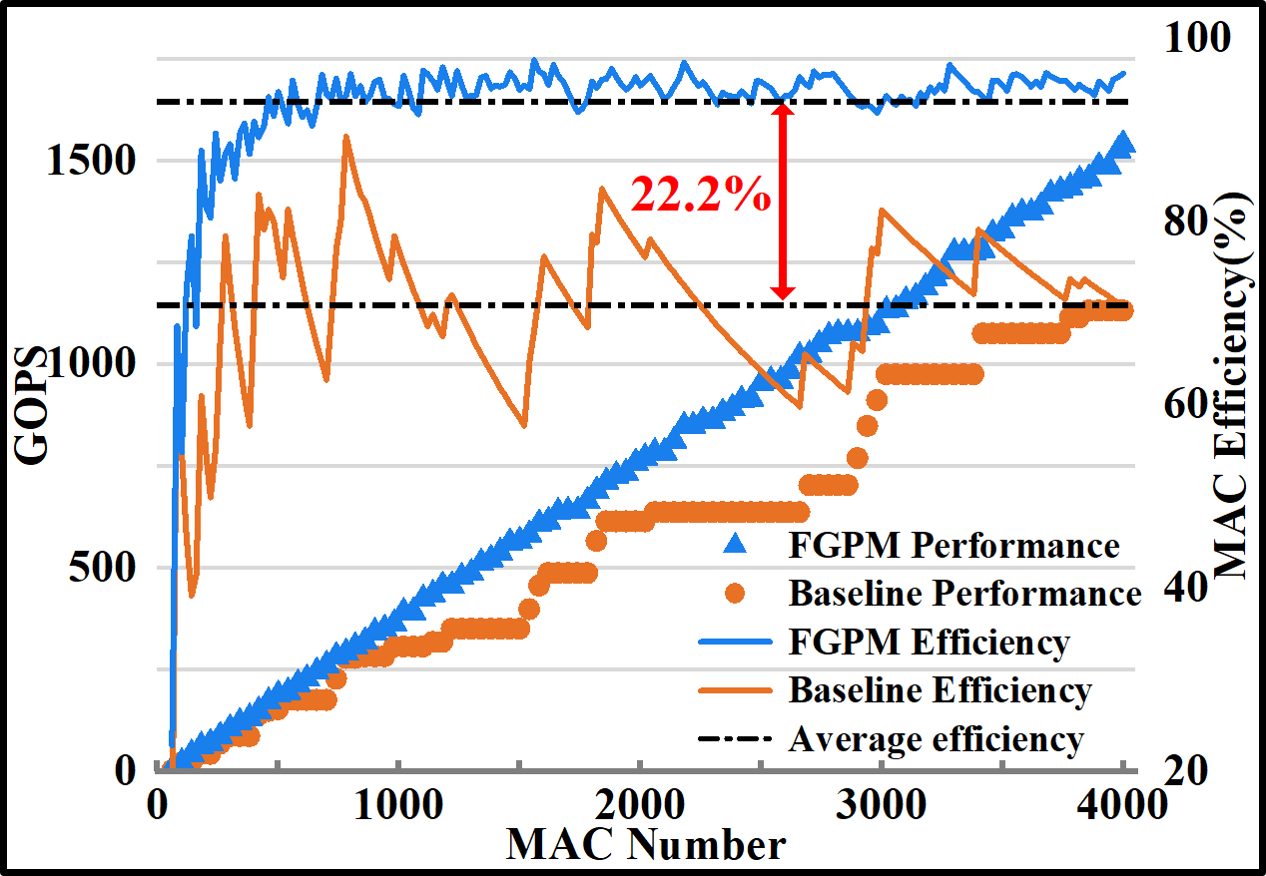}}  ~   
     \caption{Comparisons of the MAC efficiency and throughput among four LWCNNs across 60-4000 MACs run in 200 MHZ. (a) MobileNetV1; (b) MobileNetV2; (c) ShuffleNetV1; (d) ShuffleNetV2.}
    \label{fig:lwcnn operation}
    \vspace{-0.1cm} 
\end{figure*}

\begin{figure}[!t]
\centerline{\includegraphics[width=\linewidth]{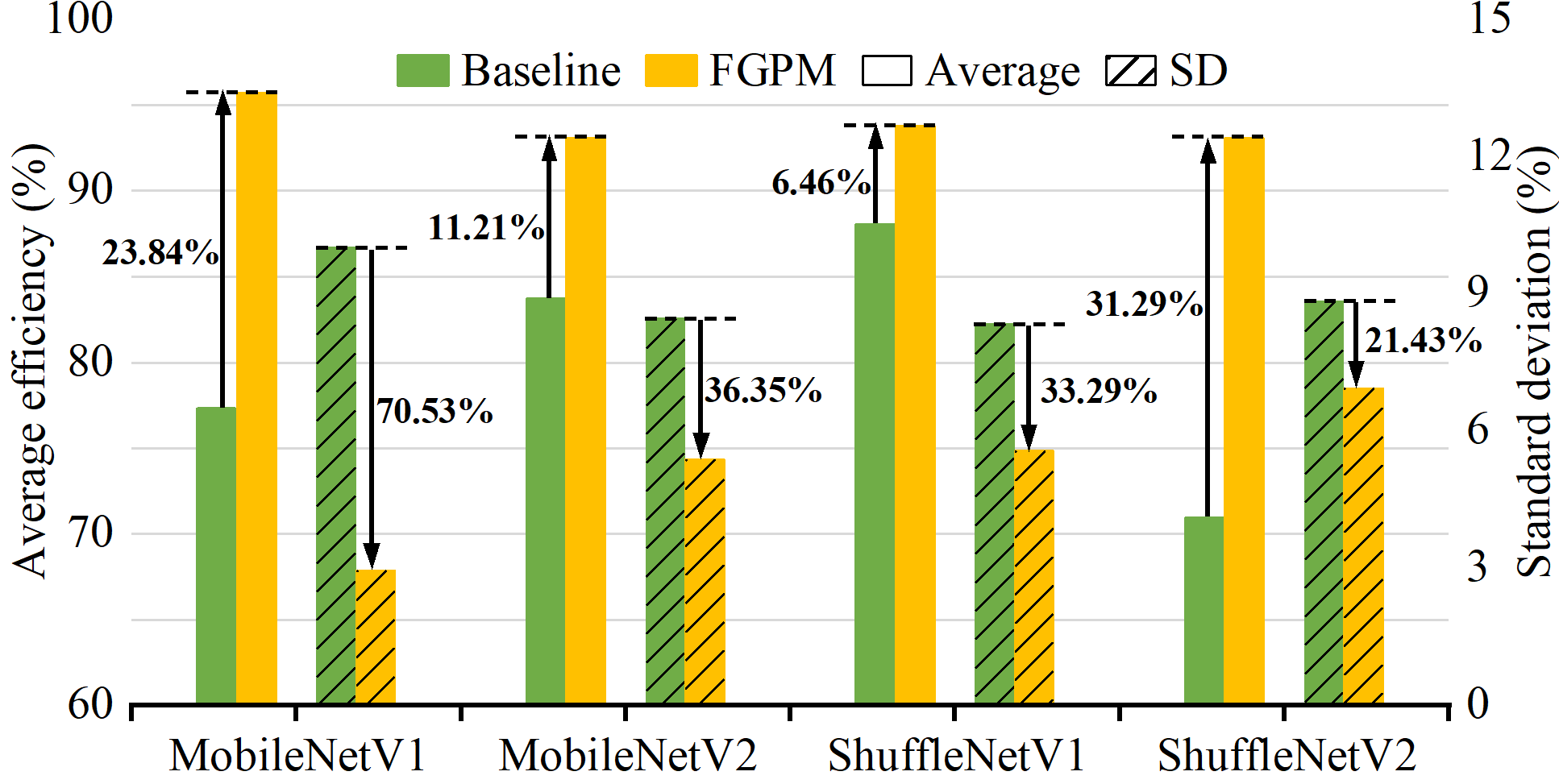}}
    \caption{Comparisons of the average efficiency and standard deviation among four LWCNNs across 60-4000 MACs.}
    \label{fig:effcmp}
\end{figure}
\begin{figure}[!t]
\centerline{\includegraphics[width=\linewidth]{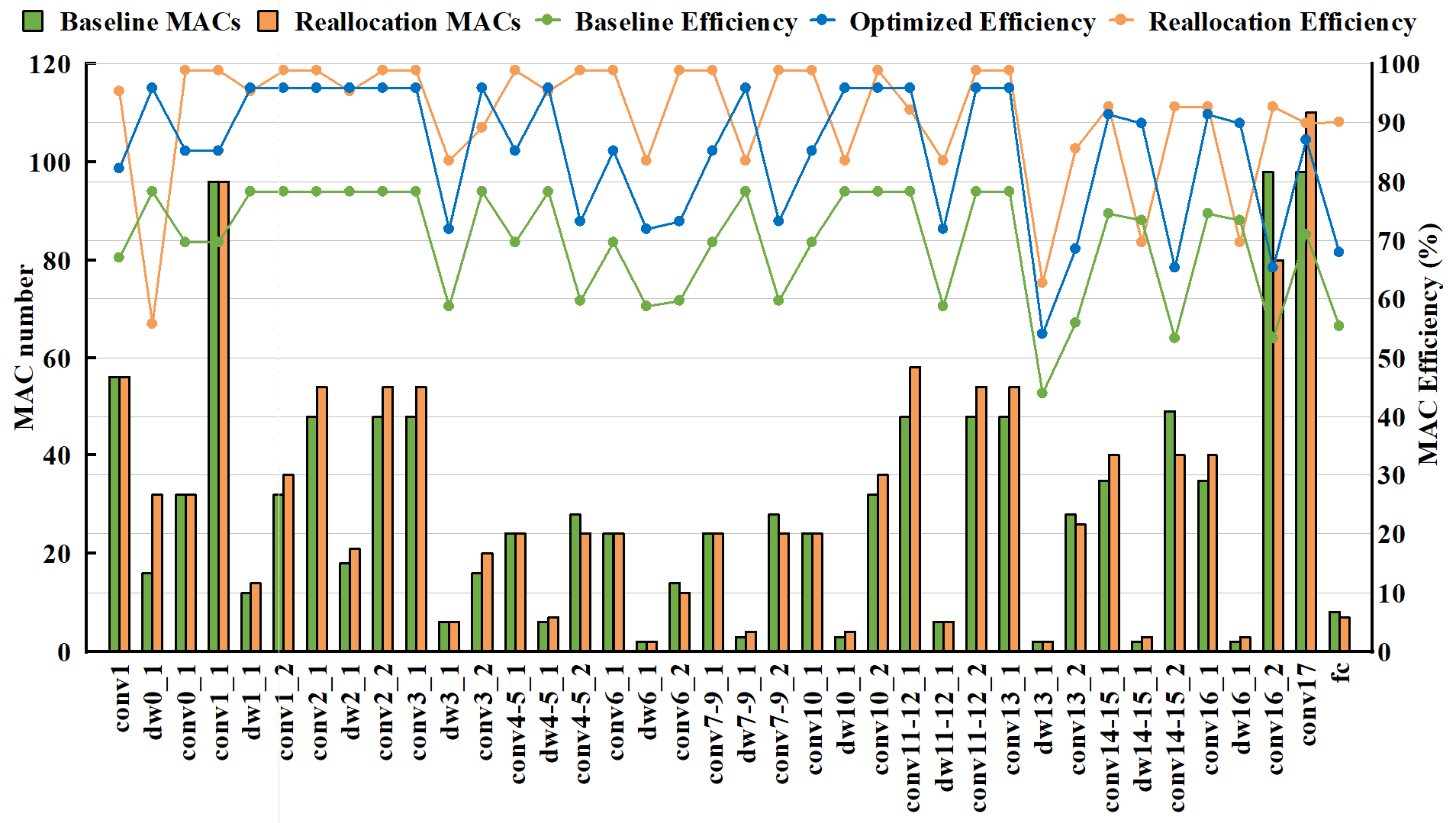}}
    \caption{MAC number and efficiency statistics of MobileNetV2 implementation with different optimization.}
    \label{fig:break}
\end{figure}

To demonstrate the effectiveness of FGPM in our multi-CE architecture, we also evaluate the enhancement in terms of theoretical throughput and scalability of FGPM compared to baseline schemes commonly employed in many streaming accelerators with the same design constraints.
Note that the evaluation does not consider the DRAM bandwidth limitations since the requirement is greatly reduced. 
The evaluation process, driven by the dynamic parallelism tuning algorithm, is shown in Fig.~\ref{fig:lwcnn operation}.
The throughput of the baseline scheme exhibits a significant staircase effect, which is caused by the fact that the increased computing resources remain underutilized and thus have no contribution to the throughput of the bottleneck layers. 
This problem is even worse in ShuffleNetV2 since most channel numbers are difficult to decompose into sufficient integer factors.
In contrast, the throughput improved by FPGM exhibits a nearly linear relationship with the growth of MAC number, indicating that the increased computing resources effectively participate in the computation in most cases, thus achieving a higher computing efficiency.
For better illustration, Fig.~\ref{fig:effcmp} unifies the average efficiency and standard deviation for configurations ranging from 60 to 4000 MACs, where the standard deviation reflects the generalization performance by measuring the dispersion of the accelerator's performance with different numbers of MACs.
The average theoretical MAC efficiency with FGPM across various MAC configurations ranges from 93.06\% to 95.68\% for four benchmark networks, representing a 6.46\% to 31.29\% improvement compared to the baseline.
Furthermore, the reduction in standard deviation also demonstrates that the proposed scheme offers better scalability for platforms with diverse computing resources.

To elaborate on the MAC efficiency improvement of the balanced dataflow strategy, the detailed implementation of MobileNetV2 with DSP number constraint from ZC706 is taken as an example for discussion.
Fig.~\ref{fig:break} shows the layer breakdown of DSP usage and MAC efficiency under different optimization schemes,
where ``baseline" means allocating MACs and processing dataflow using the original method without any optimizations, 
``optimized" alleviates the data congestion with the dataflow-oriented buffer scheme for FRCEs,
and ``reallocation" further adopts the FGPM to reallocate computing resources for each layer based on the "optimized" design.
We can see that the computational efficiency in the ``baseline" scheme is relatively low since the conventional line buffers cannot promise sufficient bandwidth for layers with padding or large strides.
Benefiting from the proposed buffer scheme, the actual MAC efficiency in the ``optimized" scheme is elevated from 69.13\% to 84.79\% compared to the baseline,
which is nearly the same as the theoretical efficiency in the current allocation scheme.
Furthermore, FGPM improves both the number of involved PEs and theoretical computational efficiency by rescheduling PEs in the ``reallocation" scheme, leading to an improvement in system throughput by 11.29\% under the same resource constraints.

\begin{table}[!t]
 \renewcommand{\arraystretch}{1.1}
\caption{Resource Utilization}
\label{tab:resource}
\resizebox{\linewidth}{!}{%
\begin{tabular}{|c|c|c|c|c|}
\hline
             & LUT             & DFF             & BRAM36K        & DSP          \\ \hline
Available    & 218600          & 437200          & 545            & 900          \\ \hline
MobileNetV2  & 163087 (74.61\%) & 189476 (43.34\%) & 329.5 (60.46\%) & 844 (93.78\%) \\ \hline
ShuffleNetV2 & 117554 (53.78\%) & 177863 (40.68\%) & 209 (38.35\%)   & 853 (94.78\%) \\ \hline
\end{tabular}%
}
\end{table}

\begin{table}[!t]
\renewcommand{\arraystretch}{1.1}
\caption{Performance Summary}
\label{tab:peformance}
\resizebox{\linewidth}{!}{%
\begin{threeparttable}
\begin{tabular}{|c|c|c|c|c|c|}
\hline
                                                               & \scriptsize MACs & \scriptsize FPS   & \begin{tabular}[c]{@{}c@{}}\scriptsize SRAM \\ \scriptsize (MB)\end{tabular} & \begin{tabular}[c]{@{}c@{}}\scriptsize Off-Chip Memory\\ \scriptsize Traffic (MB/frame)\end{tabular} & \begin{tabular}[c]{@{}c@{}}\scriptsize Latency \\ \scriptsize (ms)\end{tabular} \\ \hline
\scriptsize MobileNetV2                                                    & 1567 & 985.8 & 1.27                                                     & 2.81                                                                          & 10.63                                                   \\ \hline
\scriptsize MobileNetV2 (ZC706)  & 1569 & 981.4 & 1.75                                                     & 2.05                                                                          & 5.46                                                    \\ \hline
\scriptsize ShuffleNetV2                                                   & 1604 & 2092.4 & 0.71                                                     & 1.96                                                                          & 4.74                                                    \\ \hline
\scriptsize ShuffleNetV2 (ZC706) & 1612 & 2199.2 & 1.34                                                     & 0.98                                                                          & 1.33                                                    \\ \hline
\end{tabular}      
\end{threeparttable}
}
\end{table}

As a result, MobileNetV2 and ShuffleNetV2 are implemented on ZC706 platform, and the resource utilization is listed in Table \ref{tab:resource}.
The on-chip computing and memory resources are highly utilized thanks to the resource allocation algorithm.
Table \ref{tab:peformance} summarizes the memory occupation and overall performance under batch mode running at 200 MHZ. 
The results of the ZC706 version are also included to validate the performance impact with different group boundaries.
Note that the group boundaries are fine-tuned since the SRAM size can not actually reflect the real BRAM utilization.
With similar resource consumption and high throughput, the ZC706 versions effectively reduce DRAM access and computation latency caused by WRCE at the cost of additional SRAM consumption, demonstrating the good scalability of the proposed architecture.

\subsection{Comparisons}
\input{chapters/Ex_table0.tex}

\input{chapters/Ex_table1.tex}

The experimental results of the proposed design are compared with state-of-the-art LWCNN accelerators in terms of performance as well as computing efficiency. 
As shown in Table \ref{tab:Performance cmp},
the MobileNetV2 accelerator achieves 985.8 FPS, which outperforms \cite{wu2019high, yu2020light,  yan2021fpga,li2021dynamic,wu2021flexible}, 
but falls short of \cite{knapheide2020high, jiang2023high} with a higher number of DSPs.
Compared to \cite{ahmad2020optimizing}, which targets ShuffleNetV2 acceleration, the proposed design achieves a 6.18$\times$ throughput improvement based on the same network and platform.
To further evaluate computing resource efficiency, the metrics of \textit{Throughput/DSP} and \textit{MAC efficiency} are employed.
The former reflects the useful computations performed by each DSP on average during run-time, while the latter is calculated as the ratio of the actual throughput to the theoretical peak throughput with the same MAC number.  
The proposed accelerators achieve a \textit{Throughput/DSP} of up to 0.71 GOPS, which is 1.36 to 2.54$\times$ of reference designs with the same frequency.
In the case of \cite{jiang2023high} with a 65\% faster clock, it should be noted that the design is implemented on UltraScale+'s family FPGA (16 nm), approximately twice the size of the FPGA board used in this paper.
Benefiting from the proposed hybrid CE architecture, which eliminates off-chip access for the intermediate FMs, and the balanced dataflow strategy, which optimizes resource allocation and on-chip bandwidth,
the proposed accelerator achieves the highest MAC efficiency of 94.58\%, which can be considered as working almost at full capacity and achieves up to 5.8$\times$ improvement compared to the reference designs.
As for DSP utilization, an important but commonly ignored metric for FPGA designs,
the proposed design realizes a competitive DSP utilization of around 95\% thanks to the proposed scalable architecture and dynamic parallelism tuning algorithm.
The high DSP utilization and resource efficiency enable the proposed design to make full use of the limited computing power of the FPGA, thereby maximizing system throughput.
Additionally, the proposed design also achieves the highest energy efficiency.

Furthermore, a comparison of the MobileNetV2 accelerator with several reference designs in terms of memory occupation is presented in Table \ref{tab:mem cmp}.
The design in \cite{yu2020light} deploys a unified CE, and the intermediate FMs of each layer interact with off-chip memory, resulting in a 6.0$\times$  DRAM traffic compared to our accelerator.
The design in \cite{li2021dynamic} utilizes an adaptive row-based weight reuse scheme optimized for layer-by-layer processors, whereas our proposed work targets streaming architectures and achieves lower off-chip memory traffic and higher performance.
In \cite{yan2021fpga}, an extra buffer is deployed to store the entire input FM of SCB, allowing shortcut data to be fed directly for element-wise operation.
Meanwhile, \cite{jiang2023high} presents a streaming accelerator with a fixed data reuse scheme for all CEs. 
Although this design optimizes FM transfers with off-chip memory, it requires a large amount of SRAM for FM and weight storage.
Compared to these two designs, the proposed accelerator reduces the SRAM size by 56.67\% to 68.29\% thanks to the hybrid data reuse scheme minimizing the FM buffer size for shallow layers while avoiding the weights occupation in deep layers. 
\parskip=0pt

%% file: chapters/Ex_table0.tex
\begin{table*}[]
\centering
\renewcommand{\arraystretch}{1.3}
\resizebox{0.95\linewidth}{!}{%
\begin{threeparttable}
\caption{Performance Comparisons With Previous LWCNN Accelerators}
\label{tab:Performance cmp}
\begin{tabular}{|c|c|c|c|c|c|c|c|c|c|c|}
\hline
\textbf{}                     & \textbf{Platform}                 & \textbf{MHZ}         & \textbf{Bitwidth}  & \textbf{DSP}         & \textbf{\begin{tabular}[c]{@{}c@{}}DSP\\ Utilization\end{tabular}} & \textbf{Network} & \textbf{FPS}    & \textbf{\begin{tabular}[c]{@{}c@{}}Throughput/DSP\\ (GOPS)\end{tabular}} & \textbf{\begin{tabular}[c]{@{}c@{}}MAC\\ Efficiency\end{tabular}} & \textbf{\begin{tabular}[c]{@{}c@{}}Power Efficiency\\ (GOPS/W)\end{tabular}} \\ \hline
FPL'19\cite{wu2019high}                        & ZYNQ XCZU9EG                      & 333                  & 8                  & 2070                 & 82\%                                                               & MobileNetV2      & 809.8           & 0.23              & 17.62\%                                                           & N/A               \\ \hline
\multirow{2}{*}{FPGA’20\cite{yu2020light}}      & \multirow{2}{*}{Kintex7 XC7K325T} & \multirow{2}{*}{200} & \multirow{2}{*}{8} & \multirow{2}{*}{704} & \multirow{2}{*}{84\%}                                              & MobileNetV2      & 325.7           & 0.28              & 34.70\%                                                           & 11.40           \\ \cline{7-11} 
                              &                                   &                      &                    &                      &                                                                    & MobileNetV1      & 264.6           & 0.43              & 53.46\%                                                           & 17.57           \\ \hline
FPL'20\cite{knapheide2020high}                        & Arria10 SOC                       & 200                  & 16                 & 1220                 & 72\%                                                               & MobileNetV2      & 1050.0          & 0.52              & 64.55\%                                                           & 18.53           \\ \hline
TCASII'20 \cite{ahmad2020optimizing}                     & Virtex-7 XC7VX485T                & 200                  & 18                 & 1926                 & 68\%                                                               & ShuffleNetV1     & 787.4           & 0.11              & 28.00\%                                                           & 23.30           \\ \hline
SMC'21 \cite{fan2021hardware}                        & ZYNQ XC7Z045                      & 100                  & 8                  & N/A                    & N/A                                                                  & ShuffleNetV2     & 291.5           & N/A                 & N/A                                                                 & N/A               \\ \hline
FPL'21 \cite{yan2021fpga}                       & Virtex-7 XC7V690T                 & 150                  & 8                  & 2160                 & 60\%                                                               & MobileNetV2      & 302.3           & 0.08              & 14.00\%                                                           & 15.98           \\ \hline
TCASI'21\cite{li2021dynamic}                      & ZYNQ XCZU9EQ                      & 200                  & 8                  & 576                  & 23\%                                                               & MobileNetV2      & 381.7           & 0.40              & N/A                                                           & N/A               \\ \hline
TCAD'22 \cite{jiang2023high}                      & ZYNQ XCZU9EG                      & 333                  & 8                  & 1283                 & 51\%                                                               & MobileNetV2      & 1910.0          & 0.89              & 80.07\%                                                           & N/A               \\ \hline
TCASI'22 \cite{ShortcutFusion22}                      & AMD KCU1500                      & 200                  & 8                  & 2240                 & 41\%                                                               & EfficientNet-B1      & 213.2          & 0.15              & 19.37\%                                                           & 15.0               \\ \hline
\multirow{2}{*}{TCASI'22\cite{wu2021flexible}}     & \multirow{2}{*}{Arria10 SOC}      & \multirow{2}{*}{200} & \multirow{2}{*}{8} & \multirow{2}{*}{607} & \multirow{2}{*}{36\%}                                              & MobileNetV1      & 235.5           & 0.44              & 65.43\%                                                           & N/A               \\ \cline{7-11} 
                              &                                   &                      &                    &                      &                                                                    & MobileNetV2      & 222.2           & 0.30              & 44.46\%                                                           & N/A               \\ \hline
\multirow{2}{*}{\textbf{Our}} & \multirow{2}{*}{ZYNQ XC7Z045}     & \multirow{2}{*}{200} & \multirow{2}{*}{8} & 844                  & \textbf{94\%}                                                               & MobileNetV2      & \textbf{985.8}  & \textbf{0.70}     & \textbf{94.35\%}                                                  & \textbf{57.37}\tnote{*}  \\ \cline{5-11} 
                              &                                   &                      &                    & 853                  & \textbf{95\%}                                                               & ShuffleNetV2     & \textbf{2092.4} & \textbf{0.71}     & \textbf{94.58\%}                                                  & \textbf{61.50}\tnote{*}  \\ \hline
\end{tabular}
\begin{tablenotes}    
        \footnotesize              
        \item[*] The power data are reported by Vivado after placing \& routing.  
      \end{tablenotes}           
\end{threeparttable}
}
\vspace{-0.2cm} 
\end{table*}

%% file: chapters/Ex_table1.tex
\begin{table*}[]
\centering
\renewcommand{\arraystretch}{1.3}
\resizebox{0.75\linewidth}{!}{%
\begin{threeparttable}
\caption{Memory Comparisons With Previous MobileNetV2 Accelerators}
\label{tab:mem cmp}
\begin{tabular}{|c|c|cc|c|c|c|c|c|}
\hline
          & Platform          & \multicolumn{2}{c|}{Logic}                   & DSP  & Bitwidth & FPS    & \begin{tabular}[c]{@{}c@{}}SRAM \\ (MB) \end{tabular} & \begin{tabular}[c]{@{}c@{}}Off-Chip Memory\\ Traffic (MB/frame) \end{tabular} \\ \hline
FPGA’20\cite{yu2020light}   & Kintex7 XC7K325T  & \multicolumn{1}{c|}{LUT 173522} & DFF 241175 & 704  & 8        & 325.7  & 0.9\tnote{*}                                                      & 16.9                                                                          \\ \hline
TCASI'21\cite{li2021dynamic}  & ZYNQ XCZU9EQ      & \multicolumn{1}{c|}{LUT 125470} & DFF 143495 & 576  & 8        & 381.7  & 1.0                                                      & 3.3                                                                           \\ \hline
FPL'21 \cite{yan2021fpga}   & Virtex-7 XC7V690T & \multicolumn{1}{c|}{LUT 308449} & DFF 278926 & 2160 & 8        & 302.3  & 4.1\tnote{*}                                                      & $ \sim $3.3                                                                       \\ \hline
TCAD'22  \cite{jiang2023high}  & ZYNQ XCZU9EG      & \multicolumn{1}{c|}{LUT 170429} & DFF 154347 & 1283 & 8        & 1910.0 & 3.0\tnote{*}                                                      & $ \sim $1.4                                                                          \\ \hline
Our       & ZYNQ XC7Z045      & \multicolumn{1}{c|}{LUT 163087} & DFF 189476 & 844  & 8        & 985.8  & 1.3                                                      & 2.8                                                                          \\ \hline
\end{tabular}
\begin{tablenotes}    
        \footnotesize                
        \item[*] The SRAM size is measured by the occupation of 36Kb Block RAM and M20k for FPGAs from AMD and Intel, respectively.  
      \end{tablenotes}           
\end{threeparttable}
}
\vspace{-0.2cm} 
\end{table*}

%% file: chapters/Conclusion.tex
\section{Conclusion}

This paper proposes an efficient and scalable LWCNN accelerator along with a set of memory and computing optimizations.
Firstly, a streaming architecture with hybrid CEs is introduced based on analysis of LWCNN algorithms, effectively reducing both on-chip and off-chip memory overhead.
Secondly,
a balanced dataflow strategy is presented to promote computational efficiency by enabling a balanced workload and sufficient bandwidth for CEs.
Furthermore,
a resource-aware memory and parallelism allocation methodology is developed to fully utilize available resources while maintaining high hardware efficiency across various target platforms.
Simulation and implementation results clearly demonstrate the effectiveness of the proposed methods with a state-of-the-art performance of up to 2092.4 FPS and a remarkable PE efficiency of up to 94.58\%.


